\documentclass[twocolumn,showpacs,aps]{revtex4}

\begin{document}

 \newcommand{\bq}{\begin{equation}}
 \newcommand{\eq}{\end{equation}}
 \newcommand{\bqn}{\begin{eqnarray}}
 \newcommand{\eqn}{\end{eqnarray}}
 \newcommand{\nb}{\nonumber}
 \newcommand{\lb}{\label}
\newcommand{\PRL}{Phys. Rev. Lett.}
\newcommand{\PL}{Phys. Lett.}
\newcommand{\PR}{Phys. Rev.}
\newcommand{\CQG}{Class. Quantum Grav.}

\title{Black holes and stars  in Horava-Lifshitz theory with projectability condition}

\author{Jared Greenwald $^{a}$}
\email{Jared_Greenwald@baylor.edu}

\author{Antonios Papazoglou $^{b}$}
\email{ antonios.papazoglou@port.ac.uk}

\author{Anzhong Wang $^{a}$}
\email{anzhong_wang@baylor.edu}

\affiliation{$^{a}$ GCAP-CASPER, Physics Department, Baylor
University, Waco, TX 76798-7316, USA \\
$^{b}$  Institute of Cosmology and Gravitation, University of Portsmouth, Portsmouth PO1 3FX, UK}

\date{\today}

\begin{abstract}

We systematically study spherically symmetric static spacetimes filled with a fluid in the 
Horava-Lifshitz theory of gravity with the projectability condition, but without the detailed balance. 
We establish that when the spacetime is spatially Ricci flat the unique vacuum solution is the 
de Sitter Schwarzshcild solution, while when the spacetime has a nonzero constant curvature, 
there exist two different vacuum solutions; one is an (Einstein) static universe, and the other is 
a new spacetime. This latter spacetime is maximally symmetric and not flat. We find all the 
perfect fluid solutions for  such spacetimes, in addition to a class of anisotropic fluid solutions 
of the spatially Ricci flat spacetimes.  To construct spacetimes that represent stars,  we investigate 
junction conditions across the surfaces of stars  and obtain the general matching conditions with 
or without the presence of infinitely thin shells. It is remarkable that, in contrast to general relativity, 
the radial pressure of a star does not necessarily vanish on its surface even without the presence 
of a thin shell, due to the presence of high order derivative terms.  Applying the junction conditions 
to our explicit solutions, we show that it is possible to match smoothly these solutions (all with nonzero 
radial pressures) to vacuum spacetimes without the presence of thin matter shells on the surfaces 
of stars.
 
\end{abstract}


\maketitle

\section{Introduction}
\renewcommand{\theequation}{1.\arabic{equation}} \setcounter{equation}{0}

There has been considerable interest recently on a nonrelativistic theory of gravity proposed by Horava \cite{Horava}. Inspired by the theory of a Lifshitz scalar \cite{Lifshitz} relevant in condensed matter systems, Horava postulated a  nonrelativistic anisotropic scaling symmetry of space and time. This allows for the addition of higher spatial derivative terms in the action without their time derivative counterparts, rendering the theory power counting renormalizable. The theory is supposed to flow dynamically from a scale invariant theory in the ultraviolet (UV) to General Relativity (GR) in the  infrared (IR), thus restoring the diffeomorphism symmetry at low energies.

In the class of Horava-Lifshitz (HL) theories studied during the past year, there have been many variations regarding both the potential that is used and whether the lapse function is allowed to depend on the spatial coordinates or not. In the initial proposal by Horava \cite{Horava}, a simplifying assumption was made to reduce the number of terms in the potential. The potential was required to be derived by a ``superpotential," thus giving is it a form dubbed as detailed balance. However, it was soon realized that the breaking of detailed balance is necessary in order to obtain Minkowski vacua at low energies \cite{SVW,softbreak}. This breaking can be done minimally by adding a soft breaking term in the action, or can be done in more generality by adding all possible terms. In the present paper, we will adopt the potential of   \cite{SVW}, where all operators that conserve parity are included in the potential. Since the detailed balance potential has operators that break parity, there will be no limit of the dynamics of   the potential constructed in \cite{SVW}, unless the contributions of these operators
vanish. It can be shown that when the spacetime is static and spherically symmetric,  these operators indeed have zero contributions \cite{softbreak}.  
A second assumption that distinguishes HL theories, is whether one assumes that the laspe function $N$ is a function only of time (projectable case) or a spacetime function (nonprojectable case). In the original proposal, projectability was assumed, but since the Hamiltonian constraint is not local (it becomes an integral constraint \cite{Mukb}), it was  preferred in the literature to use the nonprojectable assumption.  However, the latter seems to be inconsistent \cite{LP} as the Poisson brackets of the theory do not form a closed structure. For this reason, in the present paper we choose to work with the projectable assumption.  
As is expected, the breaking of diffeomorphism invariance results in the existence of an extra (scalar) mode in the spectrum, which is absent in GR. This mode introduces pathologies for almost all variations of the theory, giving rise to sometimes instabilities (classical or quantum), and always to strong coupling. In particular, in the projectable case this mode exists already in Minkowski backgrounds and is either classically or quantum mechanically unstable \cite{SVW,WM}. In the nonprojectable case, this mode exists for time-dependent and spatially inhomogeneous backgrounds and  is classically unstable   \cite{BPS}. If one includes   spatial gradients of the lapse function in the nonprojectable case \cite{BPS2}, the mode can be rendered stable. The cases described above that are classically unstable can be rendered stable by higher order interactions with the price of introducing strong coupling in the theory \cite{BPS,Koz}. Strong coupling also exists   in the limit that the theory approaches GR  \cite{BPS2}, which brings the effective perturbative scale lower than the GR cutoff  \cite{PS}.

To understand the  theory further, other aspects have also been studied \cite{Others}. In particular,   solutions of this theory have  been extensively investigated in the directions of cosmology as well as black holes.  Isotropic cosmological solutions were studied in \cite{Cosmos} and revealed that the new terms added in the theory typically modify the dynamics for nonzero spatial curvature. It is interesting to note that  in \cite{Mukb} it was advocated that the projectable theory has a dark matter component built-in, due to the nonlocality of the Hamiltonian constraint. 

The other set of solutions studied so far are the spherically symmetric ones \cite{BHs}.   
 In particular, Lu, Mei and Pope found all the vacuum solutions of the HL theory with the detailed balance condition \cite{LMP}. Cai {\em et al} generalized them to topological black holes  with or without charges, and paid particular attention to their thermodynamics \cite{CCO}, while Colgain and Yavartanoo obtained a class of dyonic solutions with detailed balance and 
assuming that   high order derivative terms in the potential of the massless vector field were absent \cite{CY} (See also \cite{3K}.). Adding a linear term of the three-spatial curvature into the action, so that the detailed balance condition was softly broken, Kehagias and Sfetsos constructed a class of vacuum solutions that is asymptotically flat \cite{softbreak}. Park generalized it to the case with a nonvanishing cosmological constant \cite{Park}, while Lee, Kim and Myung studied $AdS_{2}\times S^{2}$ solutions in such a deformed generalization  of the HL theory \cite{LKM}. Capasso and Polychronakos considered the case with nontrivial lapse function and shift vector and found all the solutions { with soft breaking of the detailed balance} \cite{CP}. 
Kiritsis and Kofinas further generalized the HL theory  to include  quadratic terms of the three-dimensional spatial curvature, and found the most  general solutions \cite{KK2}, while Kiritsis himself \cite{Kiritsis} considered static spherically symmetric solutions in a version proposed  in  \cite{BPS2}. Other kinds of solutions can be found in \cite{Others,BHs}.
It should be noted that all the solutions mentioned above were constructed without assuming  the projectability condition. As we shall show in this paper, all such solutions can be written in a canonical ADM form that exhibits explicitly the
projectability condition. In addition, static spherically symmetric spacetimes with the projectability condition were studied recently by  Tang and Chen, and some solutions with   or without
 charge were found \cite{TC}.   

In this paper, we systematically study static  spherically symmetric solutions in the projectable case, with the full potential of \cite{SVW}. We first present a self-contained introduction of the theory in Sec. II, and then {  in Sec. III we write down the equations of motion for projectable gauge that we will work with.} In Secs. IV and V, we present all the vacuum solutions as well as solutions in the presence of a perfect fluid for two cases: the case that the spatial curvature vanishes (Sec. IV) and the case where the spatial curvature is a nonzero constant (Sec. V). In the former case we also present a class of solutions that represent an anisotropic fluid whose radial pressure is proportional to its tangential one.  
In Sec. VI, we first develop the general junction conditions across the surface of a star with or without the presence
of an infinitely thin shell. Then, we   join our vacuum solutions with the ones of a fluid by requiring that no thin shell be present on the surface of the star.  Finally, in Sec. VI we present our main results. Three appendices, A, B and C, are also included. In Appendix A,   the general expressions { of the $\left(F_{s}\right)_{ij}$ tensors appearing in the equations of motion} for static spherically symmetric spacetimes are given, while in Appendix B  static solutions in GR are studied in the canonical Arnowitt-Deser-Misner  (ADM) form.  { In the latter, we show explicitly  how one can bring any given static metric into an ADM form with the projectability condition. However, the kind of coordinate transformations needed to achieve this, are not allowed by the foliation-preserving diffeomorphisms of the HL theory, and  { the actions are generally not invariant.  }   In Appendix C, we calculate the trace of the extrinsic curvature $K = K^{i}_{\;\; i}$ for all the solutions found in Secs. IV and V, and
study its singular behavior.

Before proceeding, we would like to note that  spherically symmetric spacetimes in the framework of the HL theory with the projectability condition are also studied 
recently by Izumi and Mukohyama \cite{IM}, and found that, among other things,  globally static and regular perfect fluid solutions do not exist. Our results presented in this paper do not contradict to it, since in this paper we have different assumptions. In particular, the perfect fluid to be considered here usually conducts heat along the 
radial direction \cite{Santos85}, while the one considered in \cite{IM} does not. This is another peculiar feature of the HL theory: Static stars in GR do not conduct
heat. In addition, black holes in the HL theory might not black, because of the different dispersion relations \cite{IM,KK2}.

\section{Horava-Lifshitz Gravity Without Detailed Balance}

\renewcommand{\theequation}{2.\arabic{equation}} \setcounter{equation}{0}

We give a very brief introduction to HL gravity without detailed
balance, but with the projectability condition. (For further
details, see \cite{SVW,WM,WMW}.) The dynamical variables are $N, \;
N^{i}$ and $g_{ij}\; (i, \; j = 1, 2, 3)$, in terms of which the
metric takes the ADM form,
 \bq \lb{2.1}
ds^{2} = - N^{2}dt^{2} + g_{ij}\left(dx^{i} + N^{i}dt\right)
     \left(dx^{j} + N^{j}dt\right).
 \eq
The theory is invariant under the scalings
 \bqn
&& t \rightarrow {\ell}^{3} t,\;\;\; x^{i}  \rightarrow {\ell}
x^{i}\,,\nb\\ \lb{2.3}&& N \rightarrow  \ell^{-2}  N ,\;  N^{i}
\rightarrow {\ell}^{-2} N^{i},\; g_{ij} \rightarrow g_{ij}.
 \eqn
It should be noted that there was a constant term $c^{2}$ in front of the lapse function $N$ in the ADM metric used in \cite{SVW},   
so that  $N$ was  rescaling as $N \rightarrow  N$, where $c$ has dimensions $[c] = [dx/dt]$. In \cite{WM,WMW} and this paper, 
we absorb it into $N$ so that now $N$
is scaling as that given above. 
The projectability condition requires a homogeneous lapse
function:
 \bq \lb{2.3a}
N = N(t), \;\;\; N^{i} =  N^{i}\left(t, x^{k}\right),\;\;\; g_{ij}
= g_{ij} \left(t, x^{k}\right).
 \eq
The form of the metric is invariant { under  the foliation-preserving diffeomorphisms of the HL theory,}
 \bq
\lb{gauge} \tilde{t}  =  t +   \chi^{0}(t), \;\;\;
\tilde{x}^{i} = x^{i} +  \chi^{i} \left(t, x^{k}\right).
 \eq

The total action consists of kinetic, potential and matter parts,
 \bqn \lb{2.4}
S = \zeta^2\int dt d^{3}x N \sqrt{g} \left({\cal{L}}_{K} -
{\cal{L}}_{{V}}+\zeta^{-2} {\cal{L}}_{M} \right),
 \eqn
where $g={\rm det}\,g_{ij}$, and
 \bqn \lb{2.5}
{\cal{L}}_{K} &=& K_{ij}K^{ij} - \left(1-\xi\right)  K^{2},\nb\\
{\cal{L}}_{{V}} &=& 2\Lambda - R + \frac{1}{\zeta^{2}}
\left(g_{2}R^{2} +  g_{3}  R_{ij}R^{ij}\right)\nb\\
& & + \frac{1}{\zeta^{4}} \left(g_{4}R^{3} +  g_{5}  R\;
R_{ij}R^{ij}
+   g_{6}  R^{i}_{j} R^{j}_{k} R^{k}_{i} \right)\nb\\
& & + \frac{1}{\zeta^{4}} \left[g_{7}R\nabla^{2}R +  g_{8}
\left(\nabla_{i}R_{jk}\right)
\left(\nabla^{i}R^{jk}\right)\right].
 \eqn
Here $\zeta^{2} = 1/{16\pi G} $, the covariant derivatives and
Ricci and Riemann terms are all constructed from the three-metric $g_{ij}$,
while $K_{ij}$ is the extrinsic curvature,
 \bq \lb{2.6}
K_{ij} = \frac{1}{2N}\left(- \dot{g}_{ij} + \nabla_{i}N_{j} +
\nabla_{j}N_{i}\right),
 \eq
where $N_{i} = g_{ij}N^{j}$. The constants $\xi, g_{I}\,
(I=2,\dots 8)$  are coupling constants, and $\Lambda$ is the
cosmological constant. It should be noted that Horava included a
cross term $C_{ij}R^{ij}$, where $C_{ij}$ is the Cotton tensor.
This term scales as    $\ell^{5}$ and explicitly violates parity.
To restore parity, this term was excluded in \cite{SVW}.

In the IR limit, all the high order curvature terms (with
coefficients $g_I$) drop out, and the total action reduces when
$\xi = 0$ to the Einstein-Hilbert action.

Variation with respect to the lapse function, $N(t)$, yields the
Hamiltonian constraint,
 \bq \lb{eq1}
\int{ d^{3}x\sqrt{g}\left({\cal{L}}_{K} + {\cal{L}}_{{V}}\right)}
= 8\pi G \int d^{3}x {\sqrt{g}\, J^{t}},
 \eq
where
 \bq \lb{eq1a}
J^{t} = 2\left(N\frac{\delta{\cal{L}}_{M}}{\delta N} +
{\cal{L}}_{M}\right).
 \eq
Because of the projectability condition $N = N(t)$, the
Hamiltonian constraint takes  a nonlocal integral form. If one
relaxes projectability and allows $N = N\left(t, x^{i}\right)$,
then the corresponding variation with respect to $N$ will yield a
local super-Hamiltonian constraint ${\cal{L}}_{K} +
{\cal{L}}_{{V}} = 8\pi GJ^{t}$. 

Variation with respect to the shift, $N^{i}$, yields the
supermomentum constraint,
 \bq \lb{eq2}
\nabla_{j}\pi^{ij} = 8\pi G J^{i},
 \eq
where the supermomentum, $\pi^{ij} $, and matter current, $J^{i}$,
are
 \bqn \lb{eq2a}
\pi^{ij} &\equiv& \frac{\delta{\cal{L}}_{K}}{\delta\dot{g}_{ij}}
 = - K^{ij} + \left(1 - \xi\right) K g^{ij},\nb\\
J^{i} &\equiv& - N\frac{\delta{\cal{L}}_{M}}{\delta N_{i}}.
 \eqn
Varying with respect to $g_{ij}$, on the other hand, leads to the
dynamical equations,
 \bqn \lb{eq3}
&&
\frac{1}{N\sqrt{g}}\left(\sqrt{g}\pi^{ij}\right)^{\displaystyle{\cdot}}
= -2\left(K^{2}\right)^{ij}+2 \left(1 - \xi\right)K K^{ij}
\nb\\
& &~~ + \frac{1}{N}\nabla_k\left(N^k \pi^{ij} - N^i \pi^{jk} - N^j \pi^{ik} \right) \nb\\
& &~~ + \frac{1}{2} {\cal{L}}_{K}g^{ij}   + F^{ij} + 8\pi G
\tau^{ij},
 \eqn
where $\left(K^{2}\right)^{ij} \equiv K^{il}K_{l}^{j}$, and
 \bqn
\lb{eq3a} F^{ij} &\equiv&
\frac{1}{\sqrt{g}}\frac{\delta\left(-\sqrt{g}
{\cal{L}}_{V}\right)}{\delta{g}_{ij}}
 = \sum^{8}_{s=0}{g_{s} \zeta^{n_{s}}
 \left(F_{s}\right)^{ij} }.
 \eqn
The constants are given by $g_{0} = {2\Lambda}{\zeta^{-2}}$,
$g_{1} = -1$, and $n_{s} =(2, 0, -2, -2, -4, -4, -4, -4,-4)$. The
stress 3-tensor is defined as
 \bq \label{tau}
\tau^{ij} = {2\over \sqrt{g}}{\delta \left(\sqrt{g}
 {\cal{L}}_{M}\right)\over \delta{g}_{ij}},
 \eq
and the geometric 3-tensors $ \left(F_{s}\right)_{ij}$ are defined
as follows:
  \bqn \lb{eq3b}
\left(F_{0}\right)_{ij} &=& - \frac{1}{2}g_{ij},\nb\\
\left(F_{1}\right)_{ij} &=& R_{ij}- \frac{1}{2}Rg_{ij},\nb\\
\left(F_{2}\right)_{ij} &=& 2\left(R_{ij} -
\nabla_{i}\nabla_{j}\right)R
-  \frac{1}{2}g_{ij} \left(R - 4\nabla^{2}\right)R,\nb\\
\left(F_{3}\right)_{ij} &=& \nabla^{2}R_{ij} - \left(\nabla_{i}
\nabla_{j} - 3R_{ij}\right)R - 4\left(R^{2}\right)_{ij}\nb\\
& & +  \frac{1}{2}g_{ij}\left( 3 R_{kl}R^{kl} + \nabla^{2}R
- 2R^{2}\right),\nb\\
\left(F_{4}\right)_{ij} &=& 3 \left(R_{ij} -
\nabla_{i}\nabla_{j}\right)R^{2}
 -  \frac{1}{2}g_{ij}\left(R  - 6 \nabla^{2}\right)R^{2},\nb\\
 \left(F_{5}\right)_{ij} &=&  \left(R_{ij} + \nabla_{i}\nabla_{j}
 \right) \left(R_{kl}R^{kl}\right)
 + 2R\left(R^{2}\right)_{ij} \nb\\
& & + \nabla^{2}\left(RR_{ij}\right) - \nabla^{k}\left[\nabla_{i}
\left(RR_{jk}\right) +\nabla_{j}\left(RR_{ik}\right)\right]\nb\\
& &  -  \frac{1}{2}g_{ij}\left[\left(R - 2 \nabla^{2}\right)
\left(R_{kl}R^{kl}\right)\right.\nb\\
& & \left.
- 2\nabla_{k}\nabla_{l}\left(RR^{kl}\right)\right],\nb\\
\left(F_{6}\right)_{ij} &=&  3\left(R^{3}\right)_{ij}  +
\frac{3}{2}
\left[\nabla^{2}\left(R^{2}\right)_{ij} \right.\nb\\
 & & \left.
 - \nabla^{k}\left(\nabla_{i}\left(R^{2}\right)_{jk} + \nabla_{j}
 \left(R^{2}\right)_{ik}\right)\right]\nb\\
 & &    -  \frac{1}{2}g_{ij}\left[R^{k}_{l}R^{l}_{m}R^{m}_{k} -
 3\nabla_{k}\nabla_{l}\left(R^{2}\right)^{kl}\right],\nb\\
 \left(F_{7}\right)_{ij} &=&  2 \nabla_{i}\nabla_{j}
 \left(\nabla^{2}R\right) - 2\left(\nabla^{2}R\right)R_{ij}\nb\\
 & &    + \left(\nabla_{i}R\right)\left(\nabla_{j}R\right)
  -  \frac{1}{2}g_{ij}\left[\left(\nabla{R}\right)^{2} +
  4 \nabla^{4}R\right],\nb\\
\left(F_{8}\right)_{ij} &=&  \nabla^{4}R_{ij} -
\nabla_{k}\left(\nabla_{i}\nabla^{2} R^{k}_{j}
                            + \nabla_{j}\nabla^{2} R^{k}_{i}
                            \right)\nb\\
& & - \left(\nabla_{i}R^{k}_{l}\right)
\left(\nabla_{j}R^{l}_{k}\right)
       - 2 \left(\nabla^{k}R^{l}_{i}\right) \left(\nabla_{k}R_{jl}
       \right)\nb\\
& &    -  \frac{1}{2}g_{ij}\left[\left(\nabla_{k}R_{lm}\right)^{2}
        -
        2\left(\nabla_{k}\nabla_{l}\nabla^{2}R^{kl}\right)\right].
 \eqn

The matter quantities $(J^{t}, \; J^{i},\; \tau^{ij})$ satisfy the
conservation laws \cite{WM,WMW,CNPS},
 \bqn \lb{eq4a} & &
 \int d^{3}x \sqrt{g} { \left[ \dot{g}_{kl}\tau^{kl} -
 \frac{1}{\sqrt{g}}\left(\sqrt{g}J^{t}\right)^{\displaystyle{\cdot}}
  \right.  }   \nb\\
 & &  \left.  \;\;\;\;\;\;\;\;\;\;\;\;\;\;\;\;\;\; +  \frac{2N_{k}}
 {N\sqrt{g}}\left(\sqrt{g}J^{k}\right)^{\displaystyle{\cdot}}
 \right] = 0,\\
\lb{eq4b} & & \nabla^{k}\tau_{ik} -
\frac{1}{N\sqrt{g}}\left(\sqrt{g}J_{i}
\right)^{\displaystyle{\cdot}} - \frac{N_{i}}{N}\nabla_{k}J^{k} \nb\\
& & \;\;\;\;\;\;\;\;\;\;\; - \frac{J^{k}}{N}\left(\nabla_{k}N_{i}
- \nabla_{i}N_{k}\right) =
 0.
\eqn

It should be noted that 
the energy-momentum tensor   in GR is defined as,
\bq
T^{\mu\nu} = \frac{1}{\sqrt{-g^{(4)}}} \frac{\delta\left(\sqrt{-g^{(4)}}{\cal{L}}_{M}\right)}{\delta g^{(4)}_{\mu\nu}},
\eq
where $\mu, \nu = 0, 1, 2, 3$, and $g^{(4)}_{00} = -N^{2} + N^{i}N_{i}, \; g^{(4)}_{0i} = N_{i}$, and $g^{(4)}_{ij} = g_{ij}$. 
Introducing the normal vector $n_{\nu}$ to the hypersurface $ t = const$ as,
\bq
\lb{2.7}
n_{\mu} = N\delta^{t}_{\mu}, \;\;\; n^{\mu} = \frac{1}{N} \left(- 1,  + N^{i} \right),
\eq
one can decompose $T_{\mu\nu}$ as \cite{PA01},
\bqn
\lb{2.8}
\rho_{H} &\equiv& T_{\mu\nu} n^{\mu} n^{\nu},\nb\\
s_{i}  &\equiv&  -  T_{\mu\nu} h^{(4)\mu}_{i} n^{\nu},\nb\\ 
s_{ij}  &\equiv&  T_{\mu\nu} h^{(4)\mu}_{i} h^{(4)\nu}_{j},
\eqn
where $h^{(4)}_{\mu\nu}$ is the projection operator, defined as $h^{(4)}_{\mu\nu} \equiv g^{(4)}_{\mu\nu}
+ n_{\mu}n_{\nu}$. In the GR limit, one may identify $J^{t},\; J_{i}, \; \tau_{ij}$ with $- 2\rho_{H},\; s_{i},\; s_{ij}$,
respectively.

\section{ Spherically Symmetric Static Spacetimes}

\renewcommand{\theequation}{3.\arabic{equation}} \setcounter{equation}{0}

The general spherically symmetric spacetime that preserves the form of Eq. (\ref{2.1}) 
with the projectability condition is described by the metric,  
\bqn
\lb{3.1}
ds^{2} &=& - N^{2}(t) dt^{2} + e^{2\nu(t,r)}\big(dr + N^{r}(t,r)dt\big)^{2}  \nb\\ 
& & + R^{2}(t,r)d\Omega^{2},
\eqn
in the spherical coordinates $x^{i} = (r, \theta, \phi)$, where $d\Omega^{2} \equiv d\theta^{2}  
+ \sin^{2}\theta d\phi^{2}$ and $N^{i} = \left\{N^{r}, 0, 0\right\}$. 
Clearly, it is invariant under the transformations,
\bq
\lb{3.2}
t = f(t'), \;\;\; r = g(t', r'),
\eq
where $f$ and $g$ are arbitrary functions of their indicated arguments. With this gauge freedom, 
we see that, without loss of generality, we can  set
\bq
\lb{gaugeA}
N(t) = 1, \;\; R(t, r)  = r,
\eq
a gauge we  refer to as {\em the canonical ADM gauge}. From now on we shall work with this gauge.

To consider spherically symmetric static spacetimes in the HL theory with projectability, we assume that 
there exists a timelike Killing vector, $\xi^{\mu}$, along $t$, namely ${\bf \xi} = \partial_{t}$. It
can then be shown that the Killing equations, $\xi_{\mu; \nu} + \xi_{\nu; \mu} = 0$, lead to
\bq
\lb{Killing}
\nu(t,r) = \nu(r),\;\;\; N^{r}(t,r) = N^{r}(r),
\eq
for which   the metric can be finally written as
\bqn
\lb{3.1b}
ds^{2} &=& - dt^{2} + \left(e^{\mu(r)} dt + e^{\nu(r)}dr\right)^{2} 
 + r^{2}d\Omega^{2}, 
\eqn 
 where    
 \bq
 \lb{3.1c}
 \mu = \nu + \ln N^{r},\;\;\;
 N^{r} = e^{\mu - \nu}. 
 \eq


 For the  metric (\ref{3.1b}),  we find 
\bqn
\lb{3.3a}
K_{ij} &=& e^{\mu+\nu}\Big(\mu'\delta^{r}_{i}\delta^{r}_{j} + re^{-2\nu}\Omega_{ij}\Big),\nb\\
R_{ij} &=&  \frac{2\nu'}{r}\delta^{r}_{i}\delta^{r}_{j} + e^{-2\nu}\Big[r\nu' - \big(1-e^{2\nu}\big)\Big]\Omega_{ij},\nb\\
{\cal{L}}_{K} &=& e^{2(\mu-\nu)}\left[\xi\mu'^{2} - \frac{4(1-\xi)}{r}\mu' - \frac{2(1-2\xi)}{r^{2}}\right],\nb\\
{\cal{L}}_{V} &=& \sum_{s=0}^{3}{{\cal{L}}_{V}^{(s)}},
\eqn
where  a prime denotes the ordinary derivative with respect to its indicated argument,
  $\Omega_{ij} \equiv \delta^{\theta}_{i}\delta^{\theta}_{j}  + \sin^{2}\theta\delta^{\phi}_{i}\delta^{\phi}_{j}$,
  and ${\cal{L}}_{V}^{(s)}$'s are given by Eq.(\ref{A.0}). 
 Then, we find that the Hamiltonian constraint (\ref{eq1}) reads,
 \bq 
 \lb{3.3b}
\int{\left({\cal{L}}_{K} + {\cal{L}}_{{V}} - 8\pi G J^{t}\right)e^{\nu}r^{2}dr}
= 0,
 \eq 
 while the momentum constraint (\ref{eq2}) yields,
 \bqn
 \lb{3.3c}
 \xi\Big[\mu'' + \big(\mu' - \nu'\big)\mu'\Big]  &+& \frac{2}{r}\Big[\xi\mu' + \big(1-\xi\big)\nu'\Big] - \frac{2\xi}{r^{2}}\nb\\
 &=& - 8\pi G e^{2(\nu -\mu)}v,
 \eqn
 where 
 \bq
 \lb{3.3d}
 J^{i} = e^{-(\mu + \nu)}\big(v, 0, 0\big).
 \eq
 The dynamical equations (\ref{eq3}), on the other hand, yield,
 \bqn
 \lb{3.3e}
& &  \xi\Big(2\mu'' - 2\nu'\mu' +\mu'^{2}\Big)\nb\\
 & &  ~~~~~~~~~ + \frac{4}{r}\Big[\mu' + \big(1-\xi\big)\nu'\Big]  + \frac{2(1-4\xi)}{r^{2}}\nb\\
 & &  ~~~~~~~~~~~~~~~ = - 2e^{2(\nu - \mu)}\Big(e^{-2\nu}F_{rr} + 8\pi G p_{r}\Big),~~~~\\
 \lb{3.3f}
 & & 2(1-\xi)\big(\mu'' - \mu'\nu'\big) + \big(4-3\xi\big)\mu'^{2} \nb\\
 & &  ~~~~~~~~~ + \frac{2(1-2\xi)}{r}\big(2\mu' - \nu'\big) \nb\\
 & &  ~~~~~~~~~~~~~~~  = -2e^{2(\nu - \mu)}\Bigg(\frac{1}{r^{2}}F_{\theta\theta} + 8\pi G p_{\theta}\Bigg), ~~
 \eqn
 where
 \bq
 \lb{3.3g}
 \tau_{ij} = e^{2\nu}p_{r}\delta^{r}_{i}\delta^{r}_{j} + r^{2}p_{\theta}\Omega_{ij},
 \eq
 and $F_{ij}$ is given by Eqs.(\ref{eq3a}) and (\ref{A.1}).  In this paper we define a fluid with $p_{r} = p_{\theta}$
 as a perfect fluid, which in general conducts heat flow along the radial direction \cite{Santos85}. 
 
Since the spacetime is static, one can see that now the energy conservation law (\ref{eq4a}) is satisfied identically,
while the momentum conservation (\ref{eq4b}) yields,
\bq
\lb{3.3i}
v\mu' - \big(v' - p_{r}'\big) - \frac{2}{r}\big(v - p_{r} + p_{\theta}\big) = 0.
\eq
It should be noted that this equation is not independent of Eqs.  (\ref{3.3c}), (\ref{3.3e}) and (\ref{3.3f}). As a result,
one cannot use it as an additional condition to determinate the six unknown functions, $\mu,\; \nu,\;  J^{t}, \; v, \; p_{r}, 
\; p_{\theta}$.  But, since it involves only first-order derivatives, it is often very useful to use it to replace one of the three equations,
(\ref{3.3c}), (\ref{3.3e}) and (\ref{3.3f}). In summary, in the  present case we have one integral Hamiltonian constraint (\ref{3.3b}), one momentum
constraint (\ref{3.3c}), and two dynamical equations, (\ref{3.3e}) and (\ref{3.3f}), for the six unknowns. Thus, in order to determinate them uniquely, 
additional conditions are required. 

In GR,  for a  perfect fluid 
[cf. Appendix B],   one condition usually
comes from the equation of state, $p = p(\rho)$, where $\rho$ denotes the energy density of the fluid
and is related to $J^{t}$ (but not exactly equal to it \cite{WM}). However,
in  the present case, the Hamiltonian constraint is not local and cannot be used to close the system. Therefore,
to have the system closed one may take the Hamiltonian constraint (\ref{3.3b}) as a constraint    
on $J^{t}(r)$ and then use  the three remaining differential equations  (\ref{3.3c}), (\ref{3.3e}) and (\ref{3.3f}),
together with two   additional conditions, to determinate uniquely the five unknowns, $\mu,\; \nu, \; v, \; p_{r}$
and  $p_{\theta}$. In this paper, we shall adopt this strategy and obtain   one of the two conditions by specifying  
 the spatial curvature. In particular,  we will consider two cases:
(i) the spacetime is spatially Ricci-flat, and (ii) the spatial curvature is a nonzero constant. 
Certainly, one can equally choose other physical conditions to close the system. 

To relate the decomposition of  the quantities $J^{t},\; J^{i}$ and $\tau_{ij}$ defined above to the ones introduced in Eq.(\ref{2.8}), one
may introduce another spacelike unit vector, $\chi_{\mu}$, which is orthogonal to $n_{\mu}, \; \theta_{\mu}$ and
$\phi_{\mu}$, where
\bqn
\lb{3.3j}
n_{\mu} &=& \delta^{t}_{\mu}, \;\;\; n^{\mu} = - \delta_{t}^{\mu} + e^{\mu-\nu}\delta_{r}^{\mu},\nb\\
\theta_{\mu} &=& r\delta^{\theta}_{\mu},\;\;\; \phi_{\mu} = r\sin\theta \delta^{\phi}_{\mu}.
\eqn
Such a $\chi_{\mu}$ is uniquely determined as
\bq
\lb{3.3k}
\chi^{\mu} = e^{-\nu}\delta^{\mu}_{r} , \;\;\; \chi_{\mu} = e^{\mu}\delta^{t}_{\mu} + e^{\nu} \delta^{r}_{\mu}.
\eq
In terms of the four unit vectors, $n_{\mu},\; \chi_{\mu},\; \theta_{\mu}$, and $ \phi_{\mu}$, one can decompose 
the energy-momentum tensor for an anisotropic fluid with heat flow as
\bqn
\lb{3.3l}
T_{\mu\nu} &=& \rho_{H}n_{\mu}n_{\nu} + q \big(n_{\mu} \chi_{\nu} + n_{\nu} \chi_{\mu} \big)\nb\\
& & + p_{r}\chi_{\mu} \chi_{\nu}  + p_{\theta}\big(\theta_{\mu}\theta_{\nu} + \phi_{\mu}\phi_{\nu}\big),
\eqn
where $\rho_{H}, \; q,\; p_{r}$ and $p_{\theta}$ denote, respectively, the energy density, heat flow 
along radial direction, radial, and tangential pressures, measured by the observer with the four-velocity
$n_{\mu}$. 
Then, combining it with Eq.(\ref{2.8}) one can see that such a decomposition is consistent with 
 the quantities $J^{t},\; J^{i}$ and $\tau_{ij}$ defined above with
 $v=q e^{\mu}$. 
 It should be noted that the definitions of the energy density $\rho_H$, the radial pressure $p_r$ and the heat flow $q$ are different from the ones ($\rho_o$, $p_R$, $q_o$)
 given by Eq. (\ref{EMT}), which are defined   by assuming that the fluid
 is comoving with respect to the orthogonal frame (\ref{B.5}).

\section{Spatially Ricci-flat Solutions}

\renewcommand{\theequation}{4.\arabic{equation}} \setcounter{equation}{0}

Requiring that the  spacetime be spatially Ricci flat, $R_{ij} = 0$,  we find that $\nu = 0$, and  
\bq
\lb{3.4}
{\cal{L}}_{K}  = e^{2\mu}\left[\xi\mu'^{2} - \frac{4(1-\xi)}{r}\mu' - \frac{2(1-2\xi)}{r^{2}}\right].
\eq
Then, Eqs. (\ref{A.0}) and  (\ref{A.1}) yield
\bq
\lb{3.5}
{\cal{L}}_{V}  = 2\Lambda ,\;\;\; F_{ij} =  - \Lambda g_{ij}.
\eq
Inserting the above into Eqs. (\ref{3.3b}),   (\ref{3.3c}),  (\ref{3.3e}),  (\ref{3.3f}) and  (\ref{3.3i}), we
obtain, respectively,
\bqn
\lb{3.6a}
& & \int{\left({\cal{L}}_{K} + 16\pi G \rho \right)r^{2}dr} = 0,\\
\lb{3.6b}
& & \xi\big(\mu'' + \mu'^{2}\big)  + \frac{2\xi}{r}\mu'   - \frac{2\xi}{r^{2}} = - 8\pi G e^{ -2\mu}v,\\
  \lb{3.6c}
& &  \xi\big(2\mu''   +\mu'^{2}\big) + \frac{4}{r}\mu'    + \frac{2(1-4\xi)}{r^{2}} \nb\\
& & ~~~~~~~~~~~ ~~~~~~~~~~~ = - 16\pi G e^{- 2\mu}\big(p_{r} + p_{\Lambda}\big),\\
 \lb{3.6d}
 & & 2(1-\xi)\mu'' + \big(4-3\xi\big)\mu'^{2}   + \frac{4(1-2\xi)}{r}\mu'   \nb\\
 && ~~~~~~~~~~~ ~~~~~~~~~~~  = -16\pi G e^{- 2\mu} \big(p_{\theta} + p_{\Lambda}\big),\\
 \lb{3.6e}
& &  v\mu' - \big(v' - p_{r}'\big) - \frac{2}{r}\big(v - p_{r} + p_{\theta}\big) = 0,
\eqn
where
\bq
\lb{3.6f}
\rho \equiv \rho_{\Lambda} - \frac{J^{t}}{2},\;\;\;  \rho_{\Lambda} = -  p_{\Lambda}  =  \frac{\Lambda}{8\pi G}.
\eq
To study the above equations further, we consider the cases $\xi = 0$ and $\xi \not= 0$ separately.

\subsection{$\xi = 0$}

When $\xi = 0$, from Eq.(\ref{3.6b}) we find that $v = 0$. Then, from Eq. (\ref{3.4}) we obtain
\bq
\lb{3.4a}
{\cal{L}}_{K}  = -2(2r\mu' + 1)\frac{e^{2\mu}}{r^{2}},
\eq
while Eqs. (\ref{3.6c}) - (\ref{3.6e}) reduce, respectively,
to
\bqn
\lb{3.7a}
& &   {2}{r}\mu'    + {1}  = - 8\pi G r^{2}e^{- 2\mu}\big(p_{r} + p_{\Lambda}\big),\\
 \lb{3.7b}
 & & \mu'' + 2\mu'^{2}   + \frac{2}{r}\mu'    = -8\pi G e^{- 2\mu}  \big(p_{\theta} + p_{\Lambda}\big),\\
 \lb{3.7e}
& &    p_{r}' + \frac{2}{r}\big(p_{r} - p_{\theta}\big) = 0.
\eqn

\subsubsection{de Sitter Schwarzschild Solution}

If we further require the fluid be perfect, $p_{r} = p_{\theta} = p$, Eq. (\ref{3.7e}) tells us that the pressure must be
 constant.  Without loss of generality, we can absorb this constant into $p_{\Lambda}$, and then  Eqs. (\ref{3.6c}) 
 and (\ref{3.6d}) have the solution,
\bq
\lb{dS}
\mu = \frac{1}{2}\ln\left(\frac{M}{r} + \frac{\Lambda}{3}r^{2}\right).
\eq 
Inserting it into Eq. (\ref{3.4}) we find that ${\cal{L}}_{K} = - 2\Lambda$. Then,   Eq. (\ref{3.6a}) requires
$J^{t} = 0$. This is exactly the de Sitter Schwarzschild solution written in the ADM form.
It  was first noticed in the framework of the HL theory in \cite{LMP},
and rederived later by several others. When $\Lambda < 0$, it  represents the anti-de Sitter Schwarzschild solution. 

\subsubsection{Anisotropic Fluid}

When $p_{\theta} = \gamma p_{r}$, where $\gamma$ is a constant but $\gamma \not= 1$,   from Eq. (\ref{3.7e}) we find that
\bq
\lb{3.8}
p_{r} = c_{0}r^{2(\gamma - 1)},
\eq
where $c_{0}$ is an integration constant. Substituting it into Eqs. (\ref{3.7a}) and (\ref{3.7b}), for $\Lambda = 0$ we obtain
\bqn
\lb{3.9}
{2}{r}\mu'    + {1} &=& - 8c_{0} \pi G e^{- 2\mu} r^{2\gamma},\\
r^{2} \mu'' + 2r^{2}\mu'^{2}   + {2r}\mu'    &=& -8c_{0} \pi\gamma  G e^{- 2\mu} r^{2\gamma},
\eqn
from which we find that
\bqn
\lb{3.10}
\mu &=&  \frac{1}{2}\ln\left[\frac{M}{r} + \left(\frac{r}{\ell}\right)^{2\gamma}\right] + \mu_{0},\nb\\
\mu_{0} &=& - \frac{1}{2} \ln\left(- \frac{1+ 2\gamma}{8c_{0}\pi G\ell^{2\gamma}}\right).
\eqn
Clearly, when $\gamma = 1$, choosing $c_{0} = - \Lambda/(8\pi G)$ and $\ell = \sqrt{3/\Lambda}$, the above solutions reduce 
exactly to the (anti-) de Sitter Schwarzschild solution (\ref{dS}). When $\gamma \not= 1$, for the solutions to be real, we must have
\bq
\lb{3.11a}
\gamma = \cases{ < - \frac{1}{2}, & $c_{0} > 0$,\cr
 > - \frac{1}{2}, & $c_{0} < 0$,\cr}
 \eq
 for $\ell^{2\gamma} > 0$, and 
 \bq
\lb{3.11b}
\gamma = \cases{ > - \frac{1}{2}, & $c_{0} > 0$,\cr
 < - \frac{1}{2}, & $c_{0} < 0$,\cr}
 \eq
 for $\ell^{2\gamma} < 0$.
On the other hand, inserting the above solution into Eq. (\ref{3.4a}) we obtain 
\bq
\lb{3.12}
{\cal{L}}_{K} =  \frac{16\pi G c_{0}}{r^{2(1-\gamma)}}.
\eq
Setting 
\bq
\lb{3.13}
\rho(r) =  -c_{0}{r^{2(\gamma - 1)}} + \tilde{\rho}(r),
\eq
where
\bq
\lb{3.14}
\tilde{\rho}(r)  \simeq \cases{ \rho_{c} + a_{0}r^{\beta_{0}}, & $ r \rightarrow 0$,\cr
a_{\infty}r^{\beta_{\infty}}, & $ r \rightarrow \infty$,\cr}
\eq
with $\rho_{c}, \; a_{0}, \; a_{ \infty}, \; \beta_{0}$  and $ \beta_{\infty}$ being constants, 
we find that the Hamiltonian constraint (\ref{3.6a}) requires
\bq
\lb{3.15}
\beta_{0} > - 3, \;\;\; \beta_{\infty} < -3.
\eq
However, to have the center, $r = 0$, free of spacetime singularity, from Eqs. (\ref{3.8}), (\ref{3.13}) and 
(\ref{3.14}), we find  that we must assume that
\bq
\lb{3.16}
\gamma > 1, \;\;\; \beta_{0} > 0.
\eq
Eq. (\ref{3.11a}) then shows that this is possible only when $c_{0} < 0$ if $\ell^{2\gamma} > 0$. Hence, the  pressures 
become negative in this case.  However, unlike  GR [cf. Eq. (\ref{B.9})],
all the three energy conditions: weak, strong and dominant \cite{HE72}, can be satisfied by properly choosing $\tilde{\rho}(r)$, which 
now is only  subject to the global Hamiltonian constraint (\ref{3.6a}).  On the other hand, when  $\ell^{2\gamma} < 0$, the condition
(\ref{3.16}) can be satisfied for $c_{0} > 0$. Then, both the energy density and pressures can be positive. Once again, due to the
integral form of the Hamiltonian constraint, one can always choose $\rho$ properly, so that all the three energy conditions can be
satisfied. It should be noted that the above conclusions do not contradict with the results obtained in \cite{IM}, in which perfect 
fluid with some conditions between $\rho$ and $p$ was considered.

When $r \rightarrow \infty$, the pressures become infinitely large, and a spacetime singularity is indicated to exist there. One may cut 
the spacetime at a finite radius, and then join the solution to the de Sitter Schwarzschild solution, given by Eq. (\ref{dS}). We shall 
consider this issue in Sec. VI.

\subsection{$\xi \not= 0$}

When $\xi \not= 0$, to have the system (\ref{3.6b}) - (\ref{3.6d}) closed, one additional condition is required. In this paper, we shall take
it to be  $v = 0$. Certainly, other physical conditions might be equally possible. Setting $v = 0$,
we find that Eq. (\ref{3.6b}) has the general solution,
\bq
\lb{3.17}
\mu = \ln\left(a r  + \frac{b}{r^{2}}\right),
\eq
where $a$ and $b$ are integration constants. Then, for $\Lambda = 0$ Eqs.  (\ref{3.6c})  and (\ref{3.6d}) yield,
\bqn
\lb{3.18}
p_{r} &=&  \frac{3}{16\pi G}\Bigg[a^{2}\big(3\xi - 2\big) + \frac{2b^{2}}{r^{6}}\Bigg],\nb\\
p_{\theta} &=&  \frac{3}{16\pi G}\Bigg[a^{2}\big(3\xi - 2\big) - \frac{4b^{2}}{r^{6}}\Bigg],
\eqn
while Eq. (\ref{3.6e}) is satisfied identically, as one would expect. On the other hand, for the solution (\ref{3.17}), we find that
\bq
\lb{3.19}
{\cal{L}}_{K}  = - 6\Big(a^{2}  - \frac{b^{2}}{r^{6}}\Big).
\eq
Thus, setting 
\bq
\lb{3.20}
\rho = \frac{3a^{2}}{8\pi G} + \tilde{\rho}(r),
\eq
we find that the Hamiltonian constraint (\ref{3.6a}) reads,
\bq
\lb{3.21}
\int{ \tilde{\rho}(r)r^{2}dr} = 0.
\eq
From Eqs. (\ref{3.18}) and (\ref{3.19}) we can see that the spacetime is usually singular at the center, $r = 0$, unless $b = 0$.
In the latter case, we have
\bq
\lb{3.18a}
p_{r}  =  p_{\theta} =  \frac{3\big(3\xi -2\big)a^{2}}{16\pi G} = \cases{\ge 0, & $ \xi \ge 2/3$\cr
< 0, & $\xi < 2/3$. \cr} 
\eq
That is, now the fluid is a perfect fluid with constant pressures, which are non-negative for $\xi \ge 2/3$. Setting $ \tilde{\rho}(r) = 0$,
for which the Hamiltonian constraint is satisfied identically, we find that the energy density also becomes a  positive constant. 

It is interesting to note that  $p_{r} = p_{\theta} = 0$ when $\xi = 2/3$.  In other words, the fluid becomes a dust. In GR, a dust cannot have a static configuration \cite{Santos85}. But, now  the field equations involve second-order derivatives of $\mu$ (in GR only the first-order terms are involved.),  which  produce a repulsive force to hold the collapse  \cite{Mukc}. If we further choose $\tilde{\rho} = -  {3a^{2}}/{(8\pi G)}$ so that $\rho = 0$, the corresponding spacetime
becomes vacuum. 


\section{Solutions with Non-Zero Constant Curvature}

\renewcommand{\theequation}{5.\arabic{equation}} \setcounter{equation}{0}

From Eq. (\ref{3.3a}), we find that
\bq
\lb{3.22} 
R =    \frac{2e^{-2\nu}}{r^{2}}\Big[2r \nu' - \big(1-e^{2\nu}\big)\Big].
\eq
When $R$ is a constant, say, $k$,  the above equation can be cast in the form,
\bq
\lb{3.23} 
2r \nu' + \left(1 - \frac{k}{2}r^{2}\right)e^{2\nu} - 1 = 0,
\eq
which has a particular solution $\nu = 0,\; k = 0$. These are the solutions studied in the last section. Therefore, in the following
we shall consider only the case where $\nu' \not= 0$. Then, we find that
 \bq
\lb{3.24} 
 \nu = -\frac{1}{2}\ln\left(1 - \frac{k}{6}r^{2}\right),
\eq
for which we have
\bqn
\lb{3.25}
R_{ij} &=& \frac{k}{3}g_{ij}, \;\;\;
F_{ij} = {\cal{F}}_{0}g_{ij},\nb\\
{\cal{L}}_{V} &=& 2\Lambda - k + \frac{3g_{2} + g_{3}}{3\zeta^{2}}k^{2} \nb\\
& & ~~~~~~~~ ~ + \frac{9g_{4} + 3g_{5} + g_{6}}{9\zeta^{4}}k^{3},\nb\\
{\cal{L}}_{K} &=&\left(1 - \frac{k}{6} r^{2}\right)e^{2\mu}\left[\xi \mu'^{2} - \frac{4(1-\xi)}{r}\mu'\right.\nb\\
& &~~~~~~~~~~~~~~~~~~~~~~~  \left.  - \frac{2(1-2\xi)}{r^{2}}\right],
\eqn
where
\bq
\lb{3.26}
{\cal{F}}_{0} \equiv - \Lambda + \frac{k}{6} + \frac{3g_{2} + g_{3}}{18\zeta^{2}}k^{2} + \frac{9g_{4} + 3g_{5} + g_{6}}{18\zeta^{4}}k^{3}.
\eq
 To study this case further, we again consider solutions with $\xi = 0$ and $\xi \not= 0$ separately.
 
\subsection{$\xi = 0$}

When $\xi = 0$, we find that the corresponding dynamical equations, momentum   constraint
 and the conservation law can be written, respectively, as
\bqn
\lb{3.27a}
& & 4\left(\frac{1}{r} - \frac{k}{6} r\right)\mu' + 2\left(\frac{1}{r^{2}} + \frac{k}{6}\right) \nb\\
& &  ~~~~~~~~~~~~~~~~~~~~~~~~~~ = - 2e^{-2\mu}\left({\cal{F}}_{0} + 8\pi G p_{r}\right),~~~~\\
\lb{3.27b}
& & \left(1 - \frac{k}{6} r^{2}\right)\left(\mu'' + 2 \mu'^{2}\right) + \left(\frac{2}{r} - \frac{k}{2}r\right)\mu' - \frac{k}{6} \nb\\
& & ~~~~~~~~~~~~~~~~~~~~~~~~~~ = - e^{-2\mu}\left({\cal{F}}_{0} + 8\pi G p_{\theta}\right),~~\\
\lb{3.27c}
& & v = - \frac{k}{24\pi G} e^{2\mu},  \\
\lb{3.27d}
& & p_{r}' + \frac{2}{r}\left(p_{r} - p_{\theta}\right) = - \frac{ke^{2\mu}}{24\pi G r} \left(r\mu' + 2\right).
\lb{3.27e}
\eqn
 
 \subsubsection{Vacuum Solutions}
 
When the spacetime is vacuum, $v = p_{r} = p_{\theta} = J^{t} = 0$, the above equations show that we must have
\bq
\lb{3.28a}
\mu = - \infty,\;\;\; {\cal{F}}_{0} = 0,
\eq
for which we have 
\bqn
\lb{3.28b}
 N^{r} &\equiv& e^{\mu - \nu} = 0,\;\;\;  {\cal{L}}_{K} = 0,\nb\\
\Lambda &=& \frac{k}{6}   + \frac{3g_{2} + g_{3}}{18\zeta^{2}}k^{2} + \frac{9g_{4} + 3g_{5} + g_{6}}{18\zeta^{4}}k^{3},\nb\\
{\cal{L}}_{V} &=& -  \frac{2k}{3}  +   \frac{4(3g_{2} + g_{3})}{9\zeta^{2}}k^{2} \nb\\
& & ~~~~~~~ + \frac{2(9g_{4} + 3g_{5} + g_{6})}{9\zeta^{4}}k^{3}.
\eqn
The Hamiltonian constraint (\ref{3.3b}) will be satisfied identically if the coupling constants are chosen so that ${\cal{L}}_{V} = 0$. 
These solutions are (Einstein) static universe solutions for either sign of the spatial curvature $k$ \cite{TC}.

\subsubsection{Perfect Fluid}

For a perfect fluid, 
Eqs. (\ref{3.27a}) and (\ref{3.27b}) yield,
\bq
\lb{3.28}
\left(1 - \frac{k}{6} r^{2}\right)\left(\mu'' + 2 \mu'^{2}\right)  - \frac{kr}{6}\mu' - \frac{1}{r^{2}} - \frac{k}{3} = 0.
\eq
Setting 
\bq
\lb{3.29}
\mu = \frac{1}{4}\ln\left(\sqrt{\frac{k}{6}} r\right) + \frac{1}{2}\ln w(z), \;\;\; z \equiv \sqrt{1 - \frac{k}{6} r^{2}},
\eq
we find that Eq. (\ref{3.28}) can be cast in the form,
\bq
\lb{3.30}
\big(1 - z^{2}\big)w'' - 2zw' + \left[a\big(a+1\big) - \frac{b^{2}}{1 - z^{2}}\right]w = 0,
\eq
with
\bq
\lb{3.31}
a = -\frac{1}{2} + 2i,\;\; b = \frac{3}{2}.
\eq
The general solution of Eq. (\ref{3.30}) is given by
\bq
\lb{3.32}
w = c_{1} P^{b}_{a}(z) + c_{2} Q^{b}_{a}(z),
\eq
where $c_{1}$ and $c_{2}$ are the integration constants and should be chosen so that
the solution   is real. 
$P^{b}_{a}(z)$   and $Q^{b}_{a}(z)$ are, respectively, the associated Legendre 
functions of the first and second kinds \cite{AS72}. Inserting Eq. (\ref{3.32}) into
Eq. (\ref{3.29}), we find that
\bq
\lb{3.33}
\mu = \frac{1}{4}\ln(r) + \frac{1}{2} \ln\Big[c_{1} P^{b}_{a}(z) + c_{2} Q^{b}_{a}(z)\Big]
+ \mu_{0},
\eq
where $\mu_{0} \equiv [\ln(k/6)]/8$. Then, Eqs.  (\ref{3.3a}), (\ref{3.27a}) and (\ref{3.27c}) yield, 
\bqn
\lb{3.34}
& &  {\cal{L}}_{K}  = - \frac{e^{2\mu_{0}}}{r^{3/2} }\left(1 - \frac{k}{6}r^{2}\right)\Bigg[3
\Big(c_{1} P^{b}_{a}(z) + c_{2} Q^{b}_{a}(z)\Big)\nb\\
& & ~~~~~~~~ - \frac{kr^{2}}{3\sqrt{1 - \frac{k}{6} r^{2}}}\Big(c_{1} P^{'b}_{a}(z) 
                 + c_{2} Q^{'b}_{a}(z)\Big)\Bigg],\nb\\
& & p  = -  \frac{{\cal{F}}_{0}}{8\pi G} \nb\\
& & ~~~~~ - \frac{\sqrt{r}e^{2\mu_{0}}}{16\pi G }\Bigg[\left(\frac{3}{r^{2}} 
+ \frac{k}{6}\right)\Big(c_{1} P^{b}_{a}(z) + c_{2} Q^{b}_{a}(z)\Big)\nb\\
& & ~~~~~ - \frac{k}{3}\sqrt{1 - \frac{k}{6} r^{2}}\Big(c_{1} P^{'b}_{a}(z) 
                 + c_{2} Q^{'b}_{a}(z)\Big)\Bigg],\nb\\
& & v = - \frac{k\sqrt{r}e^{2\mu_{0}}}{24\pi G} \Big(c_{1} P^{b}_{a}(z) + c_{2} Q^{b}_{a}(z)\Big).               
\eqn
Note that in writing the above expressions we did not impose any conditions obtained from
the vacuum case, so that our solutions are as much applicable as possible. In particular,
by properly choosing the parameters, we can have ${\cal{F}}_{0} < 0$ so the pressure
in the center of the star is positive and the fluid satisfies all the energy conditions
[See the discussions below.]. On the other hand,
setting
\bq
\lb{3.34a}
\rho(r) = \rho_{k} + \tilde{\rho}(r) - \frac{ {\cal{L}}_{K}}{16\pi G},
\eq
we find that the Hamiltonian constraint (\ref{3.3b}) reads
\bq
\lb{3.34b}
\int{\frac{ \tilde{\rho}(r) r^{2} dr}{\sqrt{1 - \frac{k}{6} r^{2}}}} = 0,
\eq
where
\bq
\lb{3.34c}
\rho_{k} \equiv \frac{k}{16\pi G}\Bigg(1  -   \frac{3g_{2} + g_{3}}{3\zeta^{2}}k
 - \frac{9g_{4} + 3g_{5} + g_{6}}{9\zeta^{4}}k^{2} \Bigg).
\eq

To study the singular behavior of the solution near the center, we first note that \cite{AS72}
\bqn
\lb{3.35}
 P^{b}_{a}(z) &=& \frac{1}{\Gamma(1-b)}\left(\frac{z+1}{z-1}\right)^{b/2} \nb\\
 & & ~~~~~ \times F\left(- a, a+1; 1-b;\frac{1-z}{2}\right),\nb\\
 Q^{b}_{a}(z) &=& e^{ib\pi} \left(\frac{z+1}{z-1}\right)^{b/2} \nb\\
 & & ~ \times \left[\frac{1}{2}\Gamma(b)
   F\left(- a, a+1; 1-b;\frac{1-z}{2}\right) \right.\nb\\
   & & ~~~~~ + \frac{\Gamma(1+a+b)\Gamma(-b)}{2\Gamma(1+a-b)} \left(\frac{z-1}{z+1}\right)^{b}\nb\\
   & &~~~~~ \left. \times
   F\left(- a, a+1; 1+ b;\frac{1-z}{2}\right)\right],
 \eqn
 for $|1 - z| < 2$, where $F(a,b; c; z)$ denotes the hypergeometric function. 
When $|z| \ll 1$, it is given by
 \bq
 \lb{3.35a}
 F(a,b; c; z) \simeq 1 + \frac{ab}{c} z + \frac{ab(a+1)(b+1)}{c(c+1)} z^{2} + {\cal{O}}\left(z^{3}\right).
 \eq
 Thus, as $r \rightarrow 0^{+}$, we find
 \bqn
\lb{3.36}
 P^{b}_{a}(z) & \simeq& \frac{a_{1}}{r^{3/2}}\Bigg\{ 1 - \frac{17k}{48} r^{2} -  \frac{425k^{2}}{4608} r^{4}
 + {\cal{O}}\left(r^{9/2}\right)\Bigg\}, \nb\\ 
  Q^{b}_{a}(z) &\simeq& \frac{a_{2}}{r^{3/2}} \Bigg\{ 1 - \frac{17k}{48} r^{2}  + a_{3} r^{3}
 + {\cal{O}}\left(r^{5/2}\right)\Bigg\}, 
  \eqn
 where
 \bqn
 \lb{3.36b}
 a_{1} &\equiv& \frac{2^{3/2}}{\Gamma\left(-\frac{1}{2}\right)\left(-\frac{k}{6}\right)^{3/4}},\;\;\;
 a_{2} \equiv  - i\frac{\sqrt{2}\Gamma\left(\frac{3}{2}\right)}{\left(-\frac{k}{6}\right)^{3/4}},\nb\\
 a_{3} &\equiv&  \frac{\Gamma\left(-\frac{3}{2}\right)\Gamma\left(2+2i\right)}{\Gamma\left(\frac{3}{2}\right)\Gamma\left(-1+2i\right)}
 \left(- \frac{k}{24}\right)^{3/2}.
 \eqn
 Then, from Eqs. (\ref{3.34}) and (\ref{3.34a}) we find that when
 \bq
\lb{3.38}
a_{1} c_{1} + a_{2} c_{2} = 0,
\eq
all the quantities, $ {\cal{L}}_{K},\; \rho,\; v$ and $p$ are free of singularity at the center $r = 0$. In fact,  for such a choice we have
 \bqn
\lb{3.37}
& &  {\cal{L}}_{K}  \simeq - 6 c_{2}a_{2}a_{3}e^{2\mu_{0}},\;\;\; v \simeq 0,\nb\\  
& & \rho(r)  \simeq \rho_{k} + \tilde{\rho}(0),    \;\;\; p   \simeq -  \frac{{\cal{F}}_{0}}{8\pi G},  
\eqn
 as $r \rightarrow 0^{+}$. 

\subsection{$\xi \not= 0$}

When $\xi \not= 0$, the dynamical equations and the momentum constraint read, respectively,
\bqn
\lb{3.39a}
& & \xi \left(1 - \frac{k}{6} r^{2}\right)\left(2\mu'' + \mu'^{2}\right) + \left(\frac{4}{r} - \frac{(2+\xi)k}{3}r\right)\mu'\nb\\
& & ~~~~~~~~~~~~~~~~~~~
 + \frac{2(1 -4\xi)}{r^{2}} + \frac{(1 +2\xi)k}{3} \nb\\
 & & ~~~~~~~~~~~~~~~~~~~~~~~~~~ = - 2 e^{-2\mu} \left({\cal{F}}_{0} + 8\pi G p_{r}\right),\\
 \lb{3.39b}
& &   \left(1 - \frac{k}{6} r^{2}\right)\Bigg[2(1-\xi)\mu'' + (4-3\xi)\mu'^{2}\Bigg] \nb\\
& &  ~~~~~~~~~~ + \left[\frac{4(1-2\xi)}{r} - \frac{(3 -5\xi)k}{3}r\right]\mu' -  \frac{(1 -2\xi)k}{3}\nb\\
& &  ~~~~~~~~~~~~~~~~~~~~~~~~~~  = - 2 e^{-2\mu} \left({\cal{F}}_{0} + 8\pi G p_{\theta}\right),\\
 \lb{3.39c}
& &  \xi \left(1 - \frac{k}{6} r^{2}\right)\Big(\mu'' + \mu'^{2}\Big)  + \xi \left(\frac{2}{r} - \frac{k}{2}r\right)\mu' 
  \nb\\
& &  ~~~~~~~~~~~~~~~~~~~~~~~~~~ -  \frac{2\xi}{r^{2}} + \frac{k}{3}  = - 8\pi G e^{-2\mu} v.
 \eqn
 
 \subsubsection{Vacuum Solutions}
 
 In the vacuum case, one of the solutions is still given by
 \bq
 \lb{3.40}
 \mu = -\infty, \;\; \; {\cal{F}}_{0} = 0,
 \eq
 for which we have $N^{r} = 0= {\cal{L}}_{K}$. When $\mu$ is finite, from Eqs. (\ref{3.39a}) - (\ref{3.39c})
 we find that the vacuum equations can be cast in the forms,
 \bqn
 \lb{3.41a}
 & & \xi \left(1 - \frac{k}{6} r^{2}\right)\left(2\mu'' + \mu'^{2}\right) + \left(\frac{4}{r} - \frac{(2+\xi)k}{3}r\right)\mu'\nb\\
& & ~~~~~~~~
 + \frac{2(1 -4\xi)}{r^{2}} + \frac{(1 +2\xi)k}{3}=  - 2 e^{-2\mu} {\cal{F}}_{0},\\
 \lb{3.41b}
 & &   \left(1 - \frac{k}{6} r^{2}\right)\Bigg[2(1-\xi)\mu'' + (4-3\xi)\mu'^{2}\Bigg] \nb\\
& &  ~~~~~~~~~~ + \left[\frac{4(1-2\xi)}{r} - \frac{(3 -5\xi)k}{3}r\right]\mu' -  \frac{(1 -2\xi)k}{3}\nb\\
& &  ~~~~~~~~~~~~~~~~~~~~~~~~~~  = - 2 e^{-2\mu} {\cal{F}}_{0},
\eqn
and
\bq
\lb{3.42}
\big(1-x^{2}\big)\mu_{,xx} + \frac{1}{x}\big(4 - 5x^{2}\big)\mu_{,x} - \frac{3}{x^{2}} + \frac{2(2-\xi)}{\xi} = 0,
\eq
where $x \equiv \sqrt{k/6}\; r$. Eq. (\ref{3.42}) has the general solution,
\bqn
\lb{3.43}
\mu &=& \mu_{0} + \ln(r)  + \frac{3(3\xi -2)}{2\xi k r^{2}} + \frac{\sqrt{1 - \frac{k}{6} r^{2}}}{r^{3}}\left(1 + \frac{k}{3}r^{2}\right)\nb\\
& & ~~~\times \Bigg[\mu_{1} + \frac{2-3\xi}{4\xi}{\mbox{arcsin}} \left(\sqrt{\frac{k}{6}}\; r\right)\Bigg],
\eqn
where $\mu_{0}$ and $\mu_{1}$ are integration constants. Inserting the above  into Eq. (\ref{3.41a}),
we find that it is satisfied only when 
\bq
\lb{3.43a}
\mu_{1} = 0,\;\;\; \xi = \frac{2}{3},
\eq
and for which  Eq. (\ref{3.41b}) gives  $ {\cal{F}}_{0} = 0$.
That is
\bq
\lb{3.43b}
 \Lambda - \frac{k}{6} - \frac{3g_{2} + g_{3}}{18\zeta^{2}}k^{2} - \frac{9g_{4} + 3g_{5} + g_{6}}{18\zeta^{4}}k^{3} = 0. \label{l1}
 \eq
When $\mu_{1} = 0$ and $\xi = 2/3$  it can be also shown that ${\cal{L}}_{K}  = 0$. The Hamiltonian constraint is then satisfied identically,
when ${\cal{L}}_{V}  = 0$, {\it i.e.},  
\bq
\lb{3.44}
2\Lambda - k + \frac{3g_{2} + g_{3}}{3\zeta^{2}}k^{2}   + \frac{9g_{4} + 3g_{5} + g_{6}}{9\zeta^{4}}k^{3} = 0,
\eq
as one can see from Eq. (\ref{3.25}). Therefore, provided that the coupling constants $g_{n},\; (n = 2, 3, ..., 6)$ are
chosen so that Eqs. (\ref{l1}) and (\ref{3.44}) hold, the solution
\bqn
\lb{3.45}
 \nu &=& -\frac{1}{2}\ln\left(1 - \frac{k}{6}r^{2}\right),\nb\\
 \mu &=& \ln(r)  + \mu_{0},\; (\xi = 2/3),
\eqn
represents {\em the unique vacuum solution of the HL theory with maximal symmetry for any given curvature $k$
and nonzero $\xi$}. It is interesting to
note that $\xi = 2/3$ is the case where an anisotropic Weyl symmetry exists in the UV limit \cite{Horava}. Note also that the
spacetime described by the solution (\ref{3.45}) is not flat even in the sense of the 4-dimensional geometry. For example,
the corresponding 4-dimensional Ricci scalar is given by,
\bq
\lb{3.46}
R^{(4)} = 12 e^{2\mu_{0}} \left(1 - \frac{k}{4} r^{2}\right)  + k, \; (\xi = 2/3),
\eq
which shows that the spacetime is not flat even when $k = 0$. 
 
 \subsubsection{Perfect Fluid}
 
 On the other hand, for a perfect fluid Eqs. (\ref{3.39a}) and (\ref{3.39b}) yield,
 \bqn
 \lb{3.46b}
 & & \left(1 - \frac{k}{6}r^{2}\right)\Bigg[(1-2\xi)\mu'' + 2(1-\xi)\mu'^{2}\Bigg] \nb\\
 && - \Bigg(\frac{4\xi}{r} + \frac{(1-6\xi)k}{6}r\Bigg)\mu' 
- \frac{1-4\xi}{r^{2}} - \frac{k}{3} = 0. ~~~~~
 \eqn
 Setting 
 \bq
 \lb{3.47}
 \mu = \frac{1+2\xi}{4(1-\xi)}\ln(r) +  \frac{1-2\xi}{2(1-\xi)}\ln w(r),
 \eq
 we find that Eq. (\ref{3.46b}) can be cast in the form of Eq. (\ref{3.30}), but now with
\bq
\lb{3.50}
a = \frac{1-2\xi + 4\sqrt{\xi(\xi+1) - 1}}{2(2\xi -1)}, \;\;\; b = \frac{3}{2}.
\eq
Therefore, in the present case the general solution of (\ref{3.46b}) is given by
\bq
\lb{3.51}
 \mu = \frac{1+2\xi}{4(1-\xi)}\ln(r) +  \frac{1-2\xi}{2(1-\xi)}\ln\Big[c_{1}P^{a}_{b}(z) + c_{2}Q^{a}_{b}(z)\Big],
\eq
where, as previously,  $z \equiv \sqrt{1 - \frac{k}{6}r^{2}}$. Once $\mu$ is given, from Eqs. (\ref{3.39a})
and (\ref{3.39c}) we can find $p$ and $v$, which are too complicated to be written explicitly here.

To study the asymptotical behavior of the above solutions near the center, we first notice that
$P^{b}_{a}(z)$ and $Q^{b}_{a}(z)$ take the same forms as those given by Eq. (\ref{3.36}), as they
do not depend explicitly on the parameter $a$ as $r \rightarrow 0$. We find that
\bqn
\lb{3.52}
p &\simeq& - \frac{{\cal{F}}_{0}}{8\pi G} + \frac{3\xi(1-4\xi)}{64\pi G(1-\xi)^{2}}\left(\frac{a_{1}c_{1} + a_{2}c_{2}}{r^{3}}\right)^{\frac{1-2\xi}{1-\xi}},\nb\\
v &\simeq&   \frac{9\xi(1-2\xi)}{32\pi G(1-\xi)^{2}}\left(\frac{a_{1}c_{1} + a_{2}c_{2}}{r^{3}}\right)^{\frac{1-2\xi}{1-\xi}},\nb\\
{\cal{L}}_{K} &\simeq&   - \frac{3\xi(5-8\xi)}{4(1-\xi)^{2}}\left(\frac{a_{1}c_{1} + a_{2}c_{2}}{r^{3}}\right)^{\frac{1-2\xi}{1-\xi}},
\eqn
as $r \rightarrow 0$. Thus, when $-1/2 \le \xi < 1$, all these quantities are finite at the center for any given $c_{1}$ and $c_{2}$, provided
that $a_{1}c_{1} + a_{2}c_{2} \not= 0$. When $\xi \ge 1$ or $\xi < 1/2$, they diverge there unless  $c_{1}$ and $c_{2}$ are chosen such 
that $a_{1}c_{1} + a_{2}c_{2} = 0$. Therefore, in the present case  $c_{1}$ and $c_{2}$ must be chosen so that
\bq
\lb{3.53}
 a_{1}c_{1} + a_{2}c_{2} = \cases{ \not= 0, & $-1/2 \le \xi < 1$,\cr
 = 0, & otherwise.\cr}
 \eq

\section{Junction Conditions}

\renewcommand{\theequation}{6.\arabic{equation}} \setcounter{equation}{0}

Let us consider a surface $\Sigma$, defined by $ r  = r_{0}$, in the spacetime described by the
 metric (\ref{3.1b}), which divides the whole spacetime into two regions, the internal region
 $r < r_{0}$, and the external region $r > r_{0}$, denoted, respectively, by  $V^{-}$ and $V^{+}$. 
 Note that once the metric is cast in the form (\ref{3.1b}), the coordinates $t$ and
 $r$ are all uniquely defined. As a result, the coordinates used in $V^{+}$ and $V^{-}$ must
 be the same, i.e., 
 \bq
 \lb{4.0}
 \big\{x^{+\mu}\big\} = \big\{x^{-\mu}\big\} = (t, r, \theta,\phi). 
 \eq
Since the highest order of derivatives in the HL theory is six, one may require that the metric coefficients
be at least $C^{6}$; that is, their derivatives up to six-order exist and are continuous across $\Sigma$. However, 
this requirement often is too strict, and, in particular, will exclude the existence of   infinitely thin shells  \cite{Isreal}.
To relax this condition, from Eqs. (\ref{2.5})  and (\ref{eq3b}) we can see that the
quadratic terms of the highest derivatives are third-order, so we may require
that the metric coefficients be at least $C^{3}$. In this way we can avoid terms that are powers 
of Dirac $\delta$-functions, which mathematically are not well defined.  

For the spherically static spacetime, this condition is
still very strict, { since the quadratic terms of the highest derivatives now are only terms involving}
$\nu''^{2},\; \nu''\nu'''$, and $\mu'^{2}$,
as we can see from Eqs. (\ref{3.3a}) - (\ref{3.3f}), (\ref{A.0}) and (\ref{A.1}). Therefore, without loss
of generality, we shall assume that $\nu(r)$ and $\mu(r)$ are at least $C^{1}$  and $C^{0}$, respectively,
across the surface $ r = r_{0}$,  { and at least $C^{4}$ and $C^{1}$ elsewhere}.
Denoting quantities defined in $V^{+}\; (V^{-})$ by $F^{+} \; (F^{-})$, we find that $\mu$ and $\nu$
can be written as
\bq
\lb{4.1}
F(r) = F^{+}(r) H\left(x\right) + F^{-}(r)\left[1 - H\left(x\right)\right], 
\eq
where $F = (\mu, \nu),\; x \equiv r - r_{0}$ (It must noted that there is no confusion between 
 $x$ used in this section and the one used in Secs. IV and V.), 
\bqn
\lb{4.2}
{\mbox{limit}}_{r \rightarrow r_{0}^{+}}{\mu^{+}(r) } &=& {\mbox{limit}}_{r \rightarrow r_{0}^{-}}{\mu^{-}(r)},\nb\\
{\mbox{limit}}_{r \rightarrow r_{0}^{+}}{\nu^{+}(r) } &=& {\mbox{limit}}_{r \rightarrow r_{0}^{-}}{\nu^{-}(r)},\nb\\
{\mbox{limit}}_{r \rightarrow r_{0}^{+}}{\nu^{+}_{,r}(r) } &=& {\mbox{limit}}_{r \rightarrow r_{0}^{-}}{\nu^{-}_{,r}(r)},
\eqn
and $H(x)$ denotes the Heavside function, defined as
\bq
\lb{4.3}
H(x) = \cases{1, & $x > 0$,\cr
0, & $x < 0$,\cr}
\eq
which has the properties \cite{Wang90},
\bqn
\lb{4.4}
& & H^{n}(x) = H(x),\;\; \; \left[1 - H(x)\right]^{n} = \left[1 - H(x)\right],\nb\\
& & H(x)\left[1 - H(x)\right] = 0, \;\;\; H'(x) = \delta(x),
\eqn
 in the sense of distributions, where $\delta(x)$ denotes the Dirac delta function. Although the high-order derivatives
of $\mu$ and $\nu$ are not continuous across the hypersurface $r = r_{0}$, we assume that they all exist and are finite
 in the limits $r \rightarrow r^{\pm}_{0}$. Then, we find that
\bqn
\lb{4.5}
\mu' &=& \mu^{' D}, \;\; \mu'' = \mu^{'' D} + \left[\mu'\right]^{-}\delta\left(x\right),\nb\\
\nu' &=& \nu^{' D},\;\; \nu'' = \nu^{'' D},\nb\\
\nu''' &=& \nu^{''' D} +  \left[\nu''\right]^{-}\delta\left(x\right),\nb\\
\nu^{(4)} &=& \nu^{(4) D}H  +  \left[\nu'''\right]^{-}\delta\left(x\right)
 +  \left[\nu''\right]^{-}\delta'\left(x\right),\nb\\
\nu^{(5)} &=& \nu^{(5) D}  +  \left[\nu^{(4)}\right]^{-}\delta\left(x\right)+  \left[\nu'''\right]^{-}\delta'\left(x\right) \nb\\
  & &  ~~~~~~~ +  \left[\nu''\right]^{-}\delta''\left(x\right),
\eqn
where  
\bqn
\lb{4.6}
 \left[\nu^{(n)}\right]^{-} &\equiv &
 {\mbox{limit}}_{r \rightarrow r_{0}^{+}}{\nu^{+\; (n)}(r) } -  {\mbox{limit}}_{r \rightarrow r_{0}^{-}}{\nu^{-\;(n)}(r)},\nb\\
 F^{(n) D}  &\equiv& F^{+\; (n)}H + F^{-\; (n)}(1- H).
 \eqn
Inserting the above expressions into Eqs. (\ref{3.3a}) and (\ref{A.0}), we find that
\bqn
\lb{4.7}
{\cal{L}}_{K} &=& {\cal{L}}_{K}^{D},\;\;\; {\cal{L}}_{V}^{(0)} =   {\cal{L}}_{V}^{(0)\; D},\nb\\
{\cal{L}}_{V}^{(1)} &=&   {\cal{L}}_{V}^{(1)\; D},\;\;\; {\cal{L}}_{V}^{(2)} =   {\cal{L}}_{V}^{(2)\; D},\nb\\
{\cal{L}}_{V}^{(3)} &=&  {\cal{L}}_{V}^{(3)\; D}  +  {\cal{L}}_{V}^{(3)\; Im}\nb\\
&\equiv&   {\cal{L}}_{V}^{(3)\; D}  + \frac{8g_{7}e^{-6\nu}}{\zeta^{4}r^{3}} 
\Big[2r\nu' - \big(1 - e^{2\nu}\big)\Big] \nb\\
&& ~~~~~~~~~~~~~~~~~~~
 \times \left[\nu''\right]^{-}\delta\left(x\right), ~~
\eqn
while from Eq. (\ref{A.1}) we find that $\left(F_{n}\right)_{ij}$'s are given by Eq. (\ref{A.2}).
The superindex ``Im" represents the impulsive part of the quantity considered, which is usually proportional to
$\delta(x)$ and  its derivatives [cf. Eqs.(\ref{4.8}) and (\ref{A.4}).]. 
Separating the nondistributional from the distributional parts of the matter content as
\bqn
\lb{4.8}
J^{t} &=& J^{t, D} + J^{t,Im},\nb\\
v &=& v^{ D} + v^{Im},\nb\\
p_{r} &=& p_{r}^{ D} +  p_{r}^{Im},\nb\\
p_{\theta} &=& p_{\theta}^{ D} +  p_{\theta}^{Im},
\eqn 
we find that the Hamiltonian constraint (\ref{3.3b}) now reads
\bqn
\lb{4.9}
&& \int{\left({\cal{L}}_{K}^{D} + {\cal{L}}_{{V}}^{D} - 8\pi G J^{t, D}\right)e^{\nu}r^{2}dr}\nb\\
& & ~~~~
= \int{\left(8\pi G J^{t, Im} -  {\cal{L}}_{V}^{(3)\; Im}\right)e^{\nu}r^{2}dr}.
\eqn 
While the momentum constraint (\ref{3.3c}) and the dynamical equations
(\ref{3.3e}) and (\ref{3.3f}) remain the same in regions $V^{+}$ and $V^{-}$, but
across the thin shell at $r = r_{0}$, they read
\bqn
\lb{4.10a}
& & \xi [\mu']^{-} \delta(x) = - 8 \pi G e^{2(\nu - \mu)} v^{Im},\\
\lb{4.10b}
& & \xi [\mu']^{-} \delta(x) + e^{-2\mu} F^{Im}_{rr} = - 8 \pi G e^{2(\nu - \mu)} p_{r}^{Im},~~\\
\lb{4.10c}
& & \big(1-\xi\big) [\mu']^{-} \delta(x) + \frac{1}{r^{2}} e^{2(\nu -\mu)} F^{Im}_{\theta\theta} \nb\\
& & ~~~~~~~~~~~~~~~~~~~~~~~~
= - 8 \pi G e^{2(\nu - \mu)} p_{\theta}^{Im},
\eqn
where $ F^{Im}_{rr}$ and $F^{Im}_{\theta\theta}$ are given by Eq. (\ref{A.3}).  Assuming that the matter content has distributional contributions no more singular than a $\delta$-function, we see from above that in order to cancel the $\delta$-function derivative terms in $ F^{Im}_{rr}$ and $F^{Im}_{\theta\theta}$, it is sufficient that there is some tuning of the couplings as $g_8=8g_7/3$.

It is interesting to note that in the GR limits: $\xi  = 0$ and $\zeta \rightarrow \infty$, we have 
$F^{Im}_{rr} =  F^{Im}_{\theta\theta} = 0$, and Eqs. (\ref{4.10a}) - (\ref{4.10c}) reduce to
\bqn
\lb{4.12a}
& & p_{\theta}^{Im} = - \frac{e^{2(\mu -\nu)}}{8\pi G}   [\mu']^{-} \delta(x),\\
\lb{4.12b}
& &  v^{Im}  =  p_{r}^{Im} = 0,  \; (\xi = 0, \; \zeta \rightarrow \infty).~~
\eqn
Thus, in this limit the radial pressure of the infinitely thin shell always vanishes. This is consistent with the conclusion obtained 
early by Santos \cite{Santos85}. 

However, this is no longer true when $\xi \not= 0$ even in the low energy limit
where $F^{Im}_{rr} =  F^{Im}_{\theta\theta} = 0$, as can be seen from Eqs. (\ref{4.10a}) - (\ref{4.10c}). In particular,
when $\nu = 0$, we find that
\bqn
\lb{4.13a}
& & \xi [\mu']^{-} e^{2\mu} \delta(x) = - 8 \pi G  v^{Im},\\
\lb{4.13b}
& & \xi [\mu']^{-} e^{2\mu} \delta(x)   = - 8 \pi G   p_{r}^{Im},~~\\
\lb{4.13c}
& & \big(1-\xi\big) [\mu']^{-} e^{2\mu} \delta(x)  = - 8 \pi G   p_{\theta}^{Im}.
\eqn

This completes the general description of the junctions of a spherically symmetric star in the HL theory of gravity. In the 
rest of this section, we shall apply the above general formulas to the solutions found in the last sections. We first notice that
solutions with nonzero constant curvature $k$ cannot be matched with the ones with zero constant curvature. This is because
in the former the function $\nu$ cannot be zero for any given $r_{0}$. As a result, $\nu$  cannot be continuous across 
$r = r_{0}$. Therefore, only the solutions with the same curvature $k$ can be matched to each other. However, since $\xi$
is a running coupling constant, in principle $\xi$ can have different values at different energies. In particular, the spacetime
deep inside a very massive star  is expected to have a very high temperature, and one would expect that $\xi$ will have
different values in the regions  inside and outside of the star.   Thus, in the following we shall consider the possibility of matching
a fluid to a vacuum solution with different $\xi$. In addition, we shall consider only the match without an infinitely thin
shell at $r = r_{0}$, that is, we shall set 
\bq
\lb{4.13d}
\rho^{Im} = v^{Im} = p^{Im}_{r} = p^{Im}_{\theta}  = 0,
\eq
which implies that $\mu$ and $\nu$ must be at least $C^{1}$ and $C^{4}$, respectively.

\subsection{ Spatially Ricci Flat Solutions}

When the spacetime is spatially Ricci flat, in Sec. IV we showed that the de Sitter Schwarzschild solution (\ref{dS}) is the unique 
vacuum solution. Therefore, in this case the spacetime outside the star is uniquely described by this solution, 
\bq
\lb{4.14}
\mu_{+} = \frac{1}{2}\ln\left(\frac{M_{+}}{r} + \frac{\Lambda}{3}r^{2}\right),\;\;\; \nu_{+} = 0. 
\eq
Inside the star, two solutions were found, one is for $\xi = 0$ given by Eq. (\ref{3.10}) and the other is for $\xi \not= 0$ given
by Eq. (\ref{3.17}) with $b = 0$. In the following let us consider them separately.

\subsubsection{$\xi = 0$}

When $\xi = 0$, the spacetime inside the star is described by Eq. (\ref{3.10}), which now can be written as
\bq
\lb{4.15}
\mu_{-} = \frac{1}{2}\ln\left[\frac{M_{-}}{r} + \left(\frac{r}{\ell}\right)^{2\gamma}\right] + \mu_{0},\;\;\; \nu_{-} = 0. 
\eq
From the above expressions we can see that $\nu$ is analytical across $r = r_{0}$, while the condition that $\mu$ being
$C^{1}$ requires
\bqn
\lb{4.16a}
\mu_{+}(r_{0}) &=& \mu_{-}(r_{0}),\\
\lb{4.16b}
\mu_{+, r}(r_{0}) &=& \mu_{- ,r}(r_{0}).
\eqn
Inserting Eqs. (\ref{4.14}) and (\ref{4.15})  into the above conditions, we find that
\bqn
\lb{4.17a}
& & \frac{3M_{+} + \Lambda r^{3}_{0}}{M_{-} + r_{0}\left(\frac{r_{0}}{\ell}\right)^{2\gamma}}  = 3e^{2\mu_{0}},\\
\lb{4.17b}
& & \frac{3M_{+}  - 2 \Lambda r^{3}_{0}}{3M_{+} + \Lambda r^{3}_{0}} 
     = \frac{M_{-}  - 2\gamma r_{0}\left(\frac{r_{0}}{\ell}\right)^{2\gamma}} {M_{-} + r_{0}\left(\frac{r_{0}}{\ell}\right)^{2\gamma}},
\eqn
from which we obtain 
\bqn
\lb{4.17c}
 M_{+}  &=&\frac{1}{3} e^{2\mu_{0}}\Bigg[3M_{-} + 2(1-\gamma)r_{0}\left(\frac{r_{0}}{\ell}\right)^{2\gamma} \Bigg],\\
 \lb{4.17d}
    \Lambda &=& - 8\pi G c_{0} r^{2(\gamma -1)}_{0}.
\eqn
Note that the condition of Eq.(\ref{4.17d}) guarantees that the radial pressure is continuous across the surface $r =r_{0}$,
i.e., $p_{r}(r_{0}) = p_{\Lambda}$, as can be seen from Eq. (\ref{3.8}). 


\subsubsection{$\xi \not= 0$}

When $\xi \not= 0$, in Sec. IV, we found the perfect fluid solution given by Eq. (\ref{3.17}) with $b = 0$, that is,
\bq
\lb{4.19}
\mu_{-}  = \ln(ar), \;\;\; \nu_{-} = 0.
\eq
The corresponding pressure is given by 
\bq
\lb{4.20}
p = \frac{3(3\xi - 2)a^{2}}{16\pi G}. 
\eq
It is interesting to note that this solution is exactly the de Sitter solution in GR. However,
in the HL theory it corresponds to a perfect fluid with its pressure given by Eq. (\ref{4.20}). As shown explicitly in Sec. IV,
choosing $\tilde{\rho}(r) = 0$ the energy density becomes $\rho = 3a^{2}/(8\pi G)$ [cf. Eq. (\ref{3.20})], which satisfies 
all the three energy conditions for $4/9 \le \xi \le 4/3$. For such an internal  solution, the conditions (\ref{4.16a}) and 
(\ref{4.16b}) read
\bq
\lb{4.21}
M_{+} = 0, \;\;\; a = \sqrt{\frac{\Lambda}{3}}.
\eq

\subsection{Stars with Non-Zero Constant  Curvature}

When the spatial three-curvature $R$ is a nonzero constant, we found two vacuum solutions, one is given by Eq. (\ref{3.40})
with $\mu = -\infty\; (N^{r} = 0)$, and the other is    given by Eq. (\ref{3.45}) with $\xi = 2/3$. The one with $\mu = -\infty$
cannot be matched to any solution with finite $\mu$ across $r = r_{0}$. As a result,  the only possible solution that describes
the spacetime outside of the star in the present case is the one given by Eq. (\ref{3.45}), 
\bq
\lb{4.22}
\mu_{+} = \ln(r) + \mu_{0},\;\;\; \nu_{+} = - \frac{1}{2}\ln\left(1 - \frac{k}{6}r^{2}\right).
\eq
On the other hand, two perfect fluid solutions were found, one is for $\xi = 0$ given by (\ref{3.33}), and the other is for $\xi \not= 0$, 
given by Eq. (\ref{3.51}). Redefining the integration constants $c_{1}$ and $c_{2}$ appearing in Eq. (\ref{3.33}), we find that
in both cases the solutions can be written as
\bqn
\lb{4.23}
 \mu_{-} &=& \frac{1+2\xi}{4(1-\xi)}\ln(r) +  \frac{1-2\xi}{2(1-\xi)}\ln\Big[c_{1}P^{a}_{b}(z) + c_{2}Q^{a}_{b}(z)\Big],\nb\\
 \nu_{-} &=&  - \frac{1}{2}\ln\left(1 - \frac{k}{6}r^{2}\right).
 \eqn
 Clearly, in the present case $\nu$ is analytical across $r = r_{0}$, and we have $F^{Im}_{rr} = F^{Im}_{\theta\theta} = 0$.
 The conditions of Eq. (\ref{4.13d}) reduce, then, to those given by Eqs. (\ref{4.16a}) and (\ref{4.16b}). For the solutions
 given by Eqs. (\ref{4.22}) and (\ref{4.23}), those conditions read
 \bqn
 \lb{4.24a}
& &  \mu_{0} =  \frac{1-2\xi}{2(1-\xi)}\ln\frac{c_{1}P^{b}_{a}(z_{0}) + c_{2}Q^{b}_{a}(z_{0})}{r^{3/2}_{0}},\\
 \lb{4.24b}
 & & \frac{c_{1}P^{'b}_{a}(z_{0}) + c_{2}Q^{'b}_{a}(z_{0})}{c_{1}P^{b}_{a}(z_{0}) + c_{2}Q^{b}_{a}(z_{0})} = \frac{9z_{0}}{k r^{2}_{0}},
 \eqn
 where $z_{0} \equiv z(r_{0})$. Clearly, by properly choosing the free parameters involved in this model, the above equations
 will be satisfied.

 \section{Conclusions}
 

\renewcommand{\theequation}{7.\arabic{equation}} \setcounter{equation}{0}

 In this paper, we have systematically studied spherically symmetric static spacetimes filled with a fluid in the HL theory of gravity
 with projectability, but without detailed balance conditions.

After writing down the relevant field equations coupled with a fluid (including the Hamiltonian, momentum constraints, dynamical 
equations, and  conservation laws) in Sec. III, we systematically studied spatially Ricci flat spacetimes in Sec. IV,
and spacetimes with nonzero constant curvature in Sec. V, for both cases where the spacetimes are vacuum and filled
with a fluid. In particular, in Sec. IV we showed that the de Sitter Schwarzschild solution is the unique vacuum solution that 
is spatially flat. In this section, we found two classes of solutions coupled with a fluid. The first class, given by Eq. (\ref{3.10}),
represents spacetimes filled with an anisotropic fluid in which the tangential pressure is proportional to its radial pressure,
given by Eq. (\ref{3.8}). The second class is given by Eq. (\ref{3.17}) with $b = 0$, which represents a perfect fluid with 
constant pressure. This class of solutions actually describes the de Sitter space, but corresponds to a perfect fluid 
with positive energy density and pressure. This is in contrast to GR, where de Sitter space does not satisfy
 the strong energy condition \cite{HE72}. The main reason is that 
in the HL theory the Hamiltonian constraint becomes a global one, and has less constraint on the energy density. 
When $\xi = 2/3$, the pressure vanishes thus representing dust. In GR, dust cannot
have a static configuration and it  necessarily develops spacetime singularities  \cite{Santos85}. In the
HL theory,  higher order derivatives are present, and it is exactly the existence of these terms that produce
repulsive forces, which  prevent the collapse of the dust. This provides another concrete example in which  
a would-be caustic is regularized by the repulsive gravitational forces, created from the gradients of high
order derivatives of curvature \cite{Mukc}.

In Sec. V, we found that there are two different vacuum solutions for spacetimes with nonzero constant curvature. One is
an (Einstein) static universe, given by Eq. (\ref{3.28a}) or (\ref{3.40}), and the other is given
by Eq. (\ref{3.45}), which has the maximal symmetry and is not flat. 
The general
solutions for a perfect fluid was found explicitly, and are given, respectively,  by Eq. (\ref{3.33}) for $\xi = 0$, and 
Eq. (\ref{3.51}) for $\xi \not= 0$. 

To construct spacetimes that represent stars, we investigated the junction conditions across the surfaces of stars in Sec. VI,
and obtained the general junction conditions with/without infinitely thin shells. It is remarkable that, in contrast to
GR \cite{Santos85}, the radial pressure of the star does not necessarily vanish on the surface of the star, neither does the 
radial pressure of the thin shell.  This is due
to the high order derivatives of the spacetime curvature. As a result, a star can be formed much easily than that in GR. Applying 
those general formulas to the solutions found in Secs. IV and V, we showed that this is indeed the case. In particular, all the internal solutions have a nonzero radial pressure on the surface of the star but it is still possible for them to be matched smoothly to a vacuum spacetime without the presence of an infinitely thin shell on the surface.

 In Appendix B, we studied anisotropic  fluid with heat flow in general relativity with the metric written in an ADM form. Among
other things, we showed explicitly that  any given static solution written in an orthogonal form (\ref{metricA})   can be always brought
into    an ADM form (\ref{metricB}) with the projectability  condition by the coordinate   transformations   Eq. (\ref{trans}).  However, 
such coordinate transformations are not allowed by the restricted diffeomorphisms, (\ref{gauge}), of the HL theory. { In
particular, the extrinsic curvature tensor $K_{ij}$ and the 3-dimensional Ricci tensor $R_{ij}$ are no longer  tensors under the transformations
given by Eq. (\ref{trans}). As a result, any actions constructed from $K_{ij}$ and $R_{ij}$, such as the one given by (\ref{2.4}), are in general  not invariant and therefore   the transformed metric is not a solution of the projectable theory.  { In particular,   a vacuum solution is no longer vacuum in the new frame with the projectability condition \cite{CW}.}}

Finally, let us comment on another delicate issue, namely the singular behavior of the extrinsic curvature $K$ for some of the solutions discussed in this paper. We have not used it to identify spacetime singularities in the present paper, because  the singularities  given by  $K$ seem not to be  as serious as those given by other quantities, 
such as the Ricci curvature $R$, energy density $\rho$ and pressure $p$. For example, Cai and one of the present authors found in  \cite{CW} that  $K$ is singular at $r = (3M/|\Lambda|)^{1/3}$ for the anti-de Sitter Schwarzschild 
solution. This singularity is absent in general relativity  and the tidal forces and distortions felt by observers at these 
singularities are all finite. 
Therefore, it is 
not clear whether spacetimes beyond these points are extendable or not \cite{Ori}. It is exactly due to these 
considerations that we did not use   $K$ to identify spacetime singularities, although the singularities of $K$ are scalar ones and cannot be
removed by the foliation-preserving diffeomorphisms \cite{CW}. In fact, in Appendix C we showed explicitly  for which solutions  found in this paper $K$ is regular or singular at the center. Understanding the nature of singularities of $K$ is an important issue in the HL theory, and we wish to return to this problem in the near future.

~\\{\bf Acknowledgements:} 

We would like to express our gratitude to  R.-G. Cai, H. Lu,  E. Kiritsis, R. Maartens, S. Mukohyama, and D. Wands for 
valuable discussions and suggestions. We are particularly grateful  to Thomas Sotiriou for the critical reading of the manuscript 
 and his comments on it.

\section*{Appendix A:  Functions ${\cal{L}}_{V}^{(n)}$ and $\left(F_{s}\right)_{ij}$}
\renewcommand{\theequation}{A.\arabic{equation}} \setcounter{equation}{0}

 The Lagrangians ${\cal{L}}_{V}^{(n)}$'s in Eq.(\ref{3.3a}) for the static spherically symmetric spacetime (\ref{3.1b}) are given by
\bqn
\lb{A.0}
{\cal{L}}_{V}^{(0)} &=& 2\Lambda - \frac{2e^{-2\nu}}{r^{2}}\Big[2r\nu' -\big(1-e^{2\nu}\big)\Big],\nb\\
{\cal{L}}_{V}^{(1)} &=&  \frac{2e^{-4\nu}}{\zeta^{2}r^{4}}\Bigg\{2g_{2}\Big[2r\nu' -\big(1-e^{2\nu}\big)\Big]^{2}
                                    + g_{3}\Big[3r^{2}\nu'^{2}\nb\\
& & ~~~~~~~~~~  - 2r\big(1-e^{2\nu}\big)\nu' +\big(1-e^{2\nu}\big)^{2}\Big]\Bigg\},\nb\\
{\cal{L}}_{V}^{(2)} &=&  \frac{2e^{-6\nu}}{\zeta^{4}r^{6}}\Bigg\{4g_{4}\Big[2r\nu' -\big(1-e^{2\nu}\big)\Big]^{3}
                                    + 2g_{5}\Big[6r^{3}\nu'^{3}\nb\\
                              & & -7r^{2}\big(1-e^{2\nu}\big)\nu'^{2}  + 4r\big(1-e^{2\nu}\big)^{2}\nu' -  \big(1-e^{2\nu}\big)^{3}\Big]\nb\\
                               & & + g_{6}\Big[5r^{3}\nu'^{3}  -3r^{2}\big(1-e^{2\nu}\big)\nu'^{2}  + 3r\big(1-e^{2\nu}\big)^{2}\nu' \nb\\
                               & & -  \big(1-e^{2\nu}\big)^{3}\Big]\Bigg\},\nb\\
{\cal{L}}_{V}^{(3)} &=&  \frac{2e^{-6\nu}}{\zeta^{4}r^{6}}\Bigg\{4g_{7}\Bigg[2r^{4}\nu'\big(\nu''' - 7\nu'\nu'' + 6 \nu'^{3}\big)\nb\\
                             & & ~~~~ - r^{3}\Big[\big(1-e^{2\nu}\big)\nu''' - \big(9-7e^{2\nu}\big)\nu'\nu''\nb\\
                             & & ~~~~~~~~~ ~~~  + 2\big(5-3e^{2\nu}\big)\nu'^{3}\Big]\nb\\
                             & & ~~~~ - r^{2}\Big[\big(1-e^{2\nu}\big)\nu'' + 4\nu'^{2}\Big]\nb\\    
                             & & ~~~~ + r\big(1-e^{2\nu}\big)^{2}\nu' + \big(1-e^{2\nu}\big)^{2}\Bigg]\nb\\  
                         & &  + g_{8}\Bigg[3r^{4}\Big[\big(\nu'' - 4\nu'^{2}\big)\nu'' + 4 \nu'^{4}\Big]\nb\\
                             & & ~~~~ -2 r^{3}\big(\nu'' - 2\nu'^{2}\big)\nu'\nb\\ 
                             & & ~~~~ + r^{2}\Big[4\big(1-e^{2\nu}\big)\nu'' - \big(3-8e^{2\nu}\big)\nu'^{2}\Big]\nb\\    
                             & & ~~~~ + 8r\big(1-e^{2\nu}\big)\nu' + 6\big(1-e^{2\nu}\big)^{2}\Bigg]\Bigg\}.  
\eqn

The functions $\left(F_{s}\right)_{ij}$ defined by Eq.(\ref{eq3b})  are given by
\bqn
\lb{A.1}
\left(F_{0}\right)_{ij} &=& - \frac{1}{2} e^{2\nu}\delta^{r}_{i}\delta^{r}_{j} 
                                         - \frac{1}{2}  r ^{2} \Omega_{ij},\nb\\ 
\left(F_{1}\right)_{ij} &=&  \frac{1}{r^{2}} \big(1- e^{2\nu}\big)\delta^{r}_{i}\delta^{r}_{j} 
                                        -  r  e^{-2\nu} \nu' \Omega_{ij},\nb\\ 
\left(F_{2}\right)_{ij} &=& \frac{e^{-2\nu}}{r^{4}}\Big[4r^{2}\big(2\nu'' - 3\nu'^{2}\big) \nb\\
& & ~~~~~ ~~~ + \big(1- e^{2\nu}\big)\big(7 + e^{2\nu}\big)\Big]\delta^{r}_{i}\delta^{r}_{j}\nb\\
& &  + \frac{2e^{-4\nu}}{r^{2}}\Bigg[4r^{3}\big(2\nu''' - 7\nu'\nu'' + 6\nu'^{3}\big) \nb\\
& & ~~~~~~~~~~~~~ - 2r \nu'\big(7 - 3e^{2\nu}\big) \nb\\
& &  ~~~~~   -  \big(1 - e^{2\nu}\big)\big(7 + e^{2\nu}\big)\Bigg]\Omega_{ij},\nb\\
\left(F_{3}\right)_{ij} &=&  \frac{e^{-2\nu}}{r^{4}}\Big[3r^{2}\big(2\nu'' - 3\nu'^{2}\big) \nb\\
& & ~~~~~ ~~~ + \big(1- e^{2\nu}\big)\big(5 + e^{2\nu}\big)\Big]\delta^{r}_{i}\delta^{r}_{j}\nb\\
& &  + \frac{e^{-4\nu}}{r^{2}}\Bigg[3r^{3}\big(\nu''' - 7\nu'\nu'' + 6\nu'^{3}\big) \nb\\
& & ~~~~~~~~~~~~~ - 2r \nu'\big(5 - 2e^{2\nu}\big) \nb\\
& &  ~~~~~   -  \big(1 - e^{2\nu}\big)\big(5 + e^{2\nu}\big)\Bigg]\Omega_{ij},\nb\\
\left(F_{4}\right)_{ij} &=& \frac{4e^{-4\nu}}{r^{6}}\Big[16r^{3}\nu'\big(3\nu'' - 5\nu'^{2}\big) \nb\\
& & ~~~~~ ~~~ - 12r \big(1- e^{2\nu}\big)\big(2r\nu'' - 3r\nu'^{2} - 4\nu'\big) \nb\\
& & ~~~~~ ~~~ - \big(1- e^{2\nu}\big)\big(23 - 22 e^{2\nu}-  e^{4\nu}\big)\Big]\delta^{r}_{i}\delta^{r}_{j}\nb\\
& &  + \frac{4e^{-6\nu}}{r^{4}}\Bigg[24r^{4}\Big[\nu'\nu''' + \big(\nu''  - 11\nu'^{2}\big)\nu'' + 10\nu'^{4}\Big] \nb\\
& & ~~ - 4r^{3}\big(17 - 18e^{2\nu}\big)\nu'^{3} -  12r^{2}\big(15 - 11e^{2\nu}\big)\nu'^{2}\nb\\
& &  ~~ -   \big(1 - e^{2\nu}\big)\Big[12r^{3}\Big(\nu''' - 7\nu'\nu''\Big)\nb\\
& & ~~~~~~~~ - 48 r^{2}\nu'' + 3r\Big(1 + 7e^{2\nu}\Big)\nu' \nb\\
& & ~~~~~~~~~  - 2\big(1- e^{2\nu}\big)\big(23 + e^{2\nu}\big)\Big]\Bigg]\Omega_{ij},\nb\\
\left(F_{5}\right)_{ij} &=&   \frac{2e^{-4\nu}}{r^{6}}\Bigg\{12r^{4}\Big[\nu'\big(\nu''' -11 \nu'\nu'' + 10 \nu'^{3}\big) + \nu''^{2}\Big] \nb\\
& & ~~~~~~~~~~~ - 4r^{3}\Big[\big(1 - e^{2\nu}\big)\big(\nu''' - 7\nu'\nu'' + 6\nu'^{3}\big)\nb\\
& & ~~~~~~~~~~~~~~~~~~~~~ - \nu'\big(3\nu'' -2  \nu'^{2}\big)\Big]\nb\\
& & ~~~~~~~~~~~ + r^{2}\big(1 - e^{2\nu}\big)\big(2\nu'' - 15\nu'^{2}\big)\nb\\
& &  ~~~~~~~~~~~+ 4r\big(1 - e^{2\nu}\big)\big(13 - 2 e^{2\nu}\big)\nu'\nb\\
& &  ~~~~~~~~~~~ + \big(1 - e^{2\nu}\big)^{2}\big(23 -  e^{2\nu}\big)\Bigg\} \delta^{r}_{i}\delta^{r}_{j}\nb\\
& &  + \frac{2e^{-6\nu}}{r^{4}}\Bigg\{18r^{4}\Big[\nu'\nu''' + \nu''^{2}  - \nu'^{2}\big(11 \nu'' - 10\nu'^{2}\big)\Big] \nb\\
& &  ~~~~~~~~~~~~ - r^{3}\Big[7\big(1- e^{2\nu}\big)\nu'''   - \big(53 -49e^{2\nu}\big)\nu'\nu'' \nb\\
& & ~~~~~~~~~~~~~~~~~~~ + \big(45 - 42e^{2\nu}\big)\nu'^{3}\Big]\nb\\
& &  ~~~~~~~~~~~~ + r^{2}\Big[24\big(1- e^{2\nu}\big)\nu''  - \big(97 -69e^{2\nu}\big)\nu'^{2}\Big] \nb\\
& &  ~~~~~~~~~~~~ + r \big(1- e^{2\nu}\big)\big(13- 15e^{2\nu}\big)\nu' \nb\\
& &  ~~~~~~~~~~~~ + 2 \big(1- e^{2\nu}\big)^{2}\big(13 + e^{2\nu}\big)\Bigg\}\Omega_{ij}, \nb\\
\left(F_{6}\right)_{ij} &=&  \frac{e^{-4\nu}}{r^{6}}\Bigg\{10r^{3}\big(3\nu'' - 5\nu'^{2}\big)\nu'
     - 3r^{2}\Big[2\big(1-e^{2\nu}\big)\nu'' \nb\\
     & & ~~~~~~~~~ + 3 e^{2\nu}\nu'^{2}\Big]
     + 12r \big(1-e^{2\nu}\big)\nu' \nb\\
     & & ~~~~~~~~~ -   \big(1-e^{2\nu}\big)^{2} \big(14 + e^{2\nu}\big)\Bigg\}\delta^{r}_{i}\delta^{r}_{j} \nb\\
& &  + \frac{e^{-6\nu}}{r^{4}}\Bigg\{15r^{4}\Big[\nu'\nu''' + \big(\nu''  - \nu'^{2}\big)\big(\nu'' - 10\nu'^{2}\big)\Big] \nb\\
& & ~~~~~~~~~~ - r^{3}\Big[3\big(1 - e^{2\nu}\big)\nu''' -  3\big(1 - 7e^{2\nu}\big)\nu'\nu''\nb\\
& &  ~~~~~~~~~~~~~~~~~~ -   \big(25  + 18 e^{2\nu}\big) \nu'^{3}\Big]\nb\\
& &~~~~~~~~~~  + 3 r^{2}\Big[4\big(1 - e^{2\nu}\big)\nu'' -  \big(12 - 11e^{2\nu}\big)\nu'^{2}\Big]\nb\\
& &~~~~~~~~~~  + 12r\big(1 - e^{2\nu}\big)\big(2 - e^{2\nu}\big)\nu'\nb\\
& & ~~~~~~~~~~  + 2\big(1 - e^{2\nu}\big)^{2}\big(14 + e^{2\nu}\big)\Bigg\}\Omega_{ij},\nb\\ 
\left(F_{7}\right)_{ij} &=&   \frac{8e^{-4\nu}}{r^{6}}\Bigg\{r^{4}\Big[-2 \nu^{(4)} + 20 \nu'\nu''' \nb\\
& & ~~~~~~~~~~~~~~ + \big(15\nu'' - 82\nu'^{2}\big)\nu'' + 40\nu'^{4}\Big] \nb\\
& & ~~~~~~~~~ + 2r^{2}\Big[2\big(3 - e^{2\nu}\big)\nu'' - 3\big(5 - e^{2\nu}\big)\nu'^{2}\Big]\nb\\
& & ~~~~~~~~~ -8r \big(3 - e^{2\nu}\big)\nu' \nb\\
& & ~~~~~~~~~ - \big(1 - e^{2\nu}\big)\big(7 + e^{2\nu}\big)\Bigg\}\delta^{r}_{i}\delta^{r}_{j}\nb\\
& &  + \frac{8e^{-6\nu}}{r^{4}}\Bigg\{r^{5}\Big[- \nu^{(5)} + 16 \nu'\nu^{(4)} \nb\\
& & ~~~~~~~~~~~~~~~~ +  \big(25\nu''  - 101\nu'^{2}\big) \nu'''\nb\\
& & ~~~~~~~~~~~~~ -  \big(127\nu'' - 326 \nu'^{2}\big)\nu'\nu'' - 120\nu'^{5}\Big]\nb\\
& & ~~~~~ +  2r^{3}\Big[\big(3 - e^{2\nu}\big)\nu''' - \big(33 - 7e^{2\nu}\big)\nu'\nu''\nb\\
& & ~~~~~~~~~~~~~~~~  + \big(45 - 6e^{2\nu}\big)\nu'^{3}\Big]\nb\\
& & ~~~~~ -  2r^{2}\Big[4\big(3 - e^{2\nu}\big)\nu'' - \big(51 - 11e^{2\nu}\big)\nu'^{2}\Big]\nb\\
& & ~~~~~ + r\big(57 - 24e^{2\nu}- e^{4\nu}\big)\nu'\nb\\
& & ~~~~~ + 2\big(1 - e^{2\nu}\big)\big(7+ e^{2\nu}\big)\Bigg\}\Omega_{ij},\nb\\
\left(F_{8}\right)_{ij} &=& \frac{e^{-4\nu}}{r^{6}}\Bigg\{r^{4}\Big[6 \nu^{(4)} - 68 \nu'\nu''' \nb\\
& & ~~~~~~~~~~~~ - \big(59\nu'' - 358\nu'^{2}\big)\nu'' -224\nu'^{4}\Big] \nb\\
& & ~~~~~~~ + 2r^{3} \big(13\nu'' - 29\nu'^{2}\big)\nu'  \nb\\
& & ~~~~~~~ - r^{2}\Big[8 \big(5 - 2 e^{2\nu}\big)\nu'' -  7 \big(13- 4 e^{2\nu}\big)\nu'^{2}\Big] \nb\\
& & ~~~~~~~ + 16r\big(4 - e^{2\nu}\big) \nu'  \nb\\
& & ~~~~~~~ + 6\big(1 - e^{2\nu}\big)\big(1+ 3 e^{2\nu}\big)\Bigg\}\delta^{r}_{i}\delta^{r}_{j} \nb\\
& &  + \frac{e^{-6\nu}}{r^{4}}\Bigg\{3r^{5}\Big[\nu^{(5)} - 16 \nu'\nu^{(4)}  \nb\\
& & ~~~~~~~~~~~~~  - \big(25\nu'' - 101\nu'^{2}\big)\nu''' \nb\\
& & ~~~~~~~~~~~~~  + \big(127\nu'' - 326\nu'^{2}\big)\nu'\nu'' + 120 \nu'^{5}\Big] \nb\\
& & ~~~~~~~ + r^{4}\Big[\nu'\nu''' - \big(5\nu'' - 13\nu'^{2}\big)\nu'' -  14\nu'^{4}\Big] \nb\\
& & ~~~~~~~ - r^{3}\Big[2\big(7 - e^{2\nu}\big)\nu''' - 2\big(78 - 7e^{2\nu}\big)\nu'\nu''\nb\\
& & ~~~~~~~~~~~~~~~~ +2\big(107 - 6e^{2\nu}\big)\nu'^{3}\Big]\nb\\
& & ~~~~~~~ + r^{2}\Big[8\big(7 - e^{2\nu}\big)\nu'' - \big(277 - 30e^{2\nu}\big)\nu'^{2}\Big]\nb\\
& & ~~~~~~~ - 16 r\big(13 - 7e^{2\nu}\big)\nu' \nb\\
& & ~~~~~~~ - 6 \big(1 - e^{2\nu}\big)\big(11 - 3e^{2\nu}\big)\Bigg\}\Omega_{ij},\nb\\
\eqn
where we denoted  $\nu^{(n)} \equiv d^{n}\nu/dr^{n}$. Inserting  Eq. (\ref{4.5}) into the above expressions,    we find that
\bqn
\lb{A.2}
\left(F_{0}\right)_{ij} &=&\left(F_{0}\right)_{ij}^{D}, \;\;\; \left(F_{1}\right)_{ij}  = \left(F_{1}\right)_{ij}^{D},\nb\\
\left(F_{2}\right)_{ij} &=& \left(F_{2}\right)_{ij}^{D} 
  + 16re^{-4\nu} \left[\nu''\right]^{-}\delta\left(x\right)\Omega_{ij}, \nb\\ 
\left(F_{3}\right)_{ij} &=& \left(F_{3}\right)_{ij}^{D} 
  + 3re^{-4\nu} \left[\nu''\right]^{-}\delta\left(x\right)\Omega_{ij}, \nb\\ 
\left(F_{4}\right)_{ij} &=& \left(F_{4}\right)_{ij}^{D}  + \frac{48e^{-6\nu}}{r} \Big[2r\nu' - \big(1 - e^{2\nu}\big)\Big]\nb\\
& & ~~~~~~~~~~~~~~~~~~~~~
 \times \left[\nu''\right]^{-}\delta\left(x\right)\Omega_{ij}, \nb\\ 
\left(F_{5}\right)_{ij} &=& \left(F_{5}\right)_{ij}^{D}  + \frac{8e^{-4\nu}}{r^{3}}
 \Big[3r\nu' - \big(1 - e^{2\nu}\big)\Big]\nb\\
 & &  ~~~~~~~~~~~~~~~~~~~~~ \times \left[\nu''\right]^{-}\delta\left(x\right) \delta^{r}_{i}\delta^{r}_{j} \nb\\ 
 & & + \frac{2e^{-6\nu}}{r}
 \Big[18r\nu' - 7 \big(1 - e^{2\nu}\big)\Big]\nb\\
 & &  ~~~~~~~~~~~~~~~~~~~~~ \times \left[\nu''\right]^{-}\delta\left(x\right) \Omega_{ij}, \nb\\ 
\left(F_{6}\right)_{ij} &=& \left(F_{6}\right)_{ij}^{D}  + \frac{3e^{-6\nu}}{r}
 \Big[5r\nu' - \big(1 - e^{2\nu}\big)\Big]\nb\\
 & &  ~~~~~~~~~~~~~~~~~~~~~ \times \left[\nu''\right]^{-}\delta\left(x\right) \Omega_{ij}, \nb\\ 
\left(F_{7}\right)_{ij} &=& \left(F_{7}\right)_{ij}^{D}  - \frac{16e^{-4\nu}}{r^{2}}
 \Bigg\{\Big[\big[\nu^{(3)}\big]^{-} - 10\nu' \big[\nu''\big]^{-}\Big]\delta(x)\nb\\
 & & ~~~~~~~~~~~~~~~~~~~~~~~
 + \big[\nu''\big]^{-} \delta'(x)\Bigg\} \delta^{r}_{i}\delta^{r}_{j}\nb\\
& &  - \frac{8e^{-6\nu}}{r} \Bigg\{\Big[r^{2}\big[\nu^{(4)}\big]^{-} - 16r^{2}\nu' \big[\nu^{(3)}\big]^{-}\nb\\
 & & ~~~~~~~~~~~~~~  - r^{2}\big(25 \left\{\nu''\right\}^{+} - 101\nu'^{2}\big)\big[\nu''\big]^{-}\nb\\
 & & ~~~~~~~~~~~~~~  - 2\big(3 - e^{2\nu}\big)\big[\nu''\big]^{-}\Big]\delta(x)\nb\\
& & + r^{2}\Big[\big[\nu^{(3)}\big]^{-} - 16\nu' \big[\nu''\big]^{-}\Big]\delta'(x)\nb\\
& & +   r^{2}\big[\nu''\big]^{-} \delta''(x)  \Bigg\}\Omega_{ij}, \nb\\ 
\left(F_{8}\right)_{ij} &=& \left(F_{8}\right)_{ij}^{D}  + \frac{2e^{-4\nu}}{r^{2}}
 \Bigg\{\Big[3\big[\nu^{(3)}\big]^{-} - 34\nu' \big[\nu''\big]^{-}\Big]\delta(x)\nb\\
 & & ~~~~~~~~~~~~~~~~~~~~~~~
 + 3\big[\nu''\big]^{-} \delta'(x)\Bigg\} \delta^{r}_{i}\delta^{r}_{j}\nb\\
& &  + \frac{e^{-6\nu}}{r} \Bigg\{\Big[3r^{2}\big[\nu^{(4)}\big]^{-} - 48r^{2}\nu' \big[\nu^{(3)}\big]^{-}\nb\\
 & & ~~~~~~~~~~~~~~  - 3r^{2}\big(25 \left\{\nu''\right\}^{+} - 101\nu'^{2}\big)\big[\nu''\big]^{-}\nb\\
 & & ~~~~~~~~~~~~~~  + \big(r\nu' -14 + 2e^{2\nu}\big)\big[\nu''\big]^{-}\Big]\delta(x)\nb\\
& & ~~~~~~~~~~~ + 3r^{2}\Big[\big[\nu^{(3)}\big]^{-} - 16\nu' \big[\nu''\big]^{-}\Big]\delta'(x)\nb\\
& & ~~~~~~~~~~~ +  3 r^{2}\big[\nu''\big]^{-} \delta''(x)  \Bigg\}\Omega_{ij}, 
\eqn
where
\bqn
\lb{A.3}
\left\{\nu''\right\}^{+}  &\equiv&  \frac{1}{2}\left[{\mbox{limit}}_{r \rightarrow r^{+}_{0}}\nu^{+''}(r)
 + {\mbox{limit}}_{r \rightarrow r^{-}_{0}}\nu^{-''}(r)\right],\nb\\
 \left(F_{n}\right)_{ij}^{D} &\equiv& \left(F^{+}_{n}\right)_{ij}H(x) +  \left(F^{-}_{n}\right)_{ij}\big[1 -H(x)\big],
\eqn
with $x \equiv r - r_{0}$. Thus, we find that
\bqn
\lb{A.4}
 F^{Im}_{rr} &=& \frac{2e^{-4\nu}}{\zeta^{4}r^{2}}\Bigg\{\Bigg[4g_{5}\bigg(3\nu' - \frac{1}{r}\big(1 - e^{2\nu}\big)\bigg)[\nu'']^{-}\nb\\
 & & ~~~~~~~~~~~~~
 - 8g_{7} \bigg([\nu^{(3)}]^{-}  - 10 \nu' [\nu'']^{-}\bigg)\nb\\
 & & ~~~~~~~~~~~~~
 + g_{8} \bigg(3[\nu^{(3)}]^{-}  - 34 \nu' [\nu'']^{-}\bigg)\Bigg]\delta(x)\nb\\
 & &   ~~~~~~~~~~
 - \big(8g_{7} - 3g_{8}\big)  [\nu'']^{-}\delta'(x)\Bigg\},\nb\\
  F^{Im}_{\theta\theta} &=& \frac{(16g_{2} + 3g_{3}) re^{-4\nu}}{\zeta^{2}}[\nu'']^{-}\delta(x)\nb\\
 & & ~ + \frac{e^{-6\nu}}{r\zeta^{4}}\Bigg\{48g_{4} \Big[2r\nu' - \big(1 - e^{2\nu}\big)\Big]\left[\nu''\right]^{-}\nb\\
 & & ~~~~~~~~~~~~ + 2g_{5} \Big[18r\nu' - 7 \big(1 - e^{2\nu}\big)\Big] \left[\nu''\right]^{-}\nb\\
 & & ~~~~~~~~~~~~ + 3g_{6} \Big[5r\nu' - \big(1 - e^{2\nu}\big)\Big]\left[\nu''\right]^{-}\nb\\
 & & ~~~~~~~~~~~~  - 8g_{7}\Big[r^{2}\big[\nu^{(4)}\big]^{-} - 16r^{2}\nu' \big[\nu^{(3)}\big]^{-}\nb\\
 & & ~~~~~~~~~~~~~~~  - r^{2}\big(25 \left\{\nu''\right\}^{+} - 101\nu'^{2}\big)\big[\nu''\big]^{-}\nb\\
 & & ~~~~~~~~~~~~~~~  - 2\big(3 - e^{2\nu}\big)\big[\nu''\big]^{-}\Big] \nb\\
  & & ~~~~~~~~~~~~  + g_{8}\Big[3r^{2}\big[\nu^{(4)}\big]^{-} - 48r^{2}\nu' \big[\nu^{(3)}\big]^{-}\nb\\
 & & ~~~~~~~~~~~~~~~  - 3r^{2}\big(25 \left\{\nu''\right\}^{+} - 101\nu'^{2}\big)\big[\nu''\big]^{-}\nb\\
 & & ~~~~~~~~~~~~~~~  + \big(r\nu' -14 + 2e^{2\nu}\big)\big[\nu''\big]^{-}\Big]\Bigg\}\delta(x)\nb\\
 & & ~ -  \frac{(8g_{7} - 3g_{8})r}{\zeta^{4}e^{6\nu}}\Bigg(\big[\nu^{(3)}\big]^{-} - 16\nu' \big[\nu''\big]^{-}\Bigg)\delta'(x)\nb\\
 & & ~ -  \frac{(8g_{7} - 3g_{8})r}{\zeta^{4}e^{6\nu}}  \big[\nu''\big]^{-} \delta''(x).
  \eqn

\section*{Appendix B:  Spherically symmetric and static spacetimes in General Relativity}
\renewcommand{\theequation}{B.\arabic{equation}} \setcounter{equation}{0}

{ In this Appendix we will see how static spherically symmetric metrics in theories with unbroken diffeomorphism invariance can 
always be brought to a projectable form. Furthermore, we specialise our discussion to General Relativity and 
see how the equations of motion read in the new gauge.}

The  metric for spacetimes with  spherical symmetry takes the general form, 
\bq
\lb{B.1}
ds^{2} = g_{ab}dx^{a}dx^{b} + R^{2}d\Omega^{2},
\eq
where $a, b = 0, 1$, and $g_{ab}$ and $R$ are all functions of $x^{0}$ and $x^{1}$, and
$d\Omega^{2} \equiv d\theta^{2} + \sin^{2}\theta d\phi^{2}$. The  four-velocity of { a fluid moving radially} in such a 
spacetime  usually has only two nonvanishing components, 
\bq
\lb{B.2}
u_{\mu} = \left(u_{0}, u_{1}, 0, 0\right),  \;\; (\mu = 0, 1, 2, 3)
\eq
subject to the condition,
\bq
\lb{B.3}
u_{\lambda}u^{\lambda} = -1.
\eq
Clearly, the metric (\ref{B.1}) is invariant under the coordinate transformations,
\bq
\lb{B.4}
x^{0} = f\left({x'}^{0},  {x'}^{1}\right), \;\; x^{1} = g\left({x'}^{0},  {x'}^{1}\right),
\eq
where $f$ and $g$ are arbitrary functions of their indicated arguments. Using one degree of
the freedom, one usually sets $g_{01} = 0$. When one considers a fluid, one often uses the
other degree of freedom to choose the coordinates to be comoving with the fluid, so that the 
four-velocity of the fluid is given by $u_{\mu} \propto \delta_{\mu}^{0}$. Then, in this gauge we have
\bq
\lb{B.5}
ds^{2} =  - e^{2\Psi(\tau,r)}d\tau^{2}  + e^{2\Phi(\tau,r)}dr^{2} + R^{2}(\tau, r)d\Omega^{2},
\eq
with
\bq
\lb{B.6}
u_{\mu} = e^{\Psi}\delta_{\mu}^{\tau}. 
\eq
An anisotropic fluid with heat moving along the radial direction takes the form,
\bqn
\lb{EMT}
T_{\mu\nu} &=& \rho_o u_{\mu}u_{\nu} + p_{R}r_{\mu}r_{\nu} 
+ p_{\theta}\left(\theta_{\mu} \theta_{\nu} + \phi_{\mu} \phi_{\nu}\right)\nb\\
& & + q_o \big(u_{\mu} r_{\nu} + u_{\nu}r_{\mu}\big),
\eqn
where $r_{\mu},\; \theta_{\mu}$ and $ \phi_{\mu}$ are unit vectors, defined
by
\bq
\lb{Vectors}
r_{\mu} = e^{\Phi}\delta_{\mu}^{r}, \;\;
\theta_{\mu} = R\delta_{\mu}^{\theta}, \;\;
\phi_{\mu} = R\sin\theta \delta_{\mu}^{\phi}.
\eq
$\rho_o, \; p_{R}, \; p_{\theta}$ and $q_o$ are, respectively, the energy density, radial pressure, tangential pressure, and
heat of the fluid comoving in the orthonormal frame. Note that   the metric (\ref{B.5}) is still invariant under the rescaling,
\bq
\lb{B.7}
\tau = \tilde{f}(\tau'), \;\;\; r =  \tilde{g}(r'),
\eq
where $\tilde{f}$ and $\tilde{g}$ are arbitrary functions of their indicated arguments. 

When the spacetime is static, 
$\Psi,\; \Phi$ and $R$ become functions of $r$ only. Then, using the remaining gauge freedom (\ref{B.7})
we can always set $R(r) = r$,  so that the metric finally reads
\bq
\lb{metricA}
ds^{2} =  - e^{2\Psi(r)}d\tau^{2}  + e^{2\Phi(r)}dr^{2} + r^{2}d\Omega^{2}.
\eq

 Let us now make the coordinate transformations,
\bq
\lb{trans}
\tau = t -     \int^{r}{ \sqrt{e^{-2\Psi} - 1}\; e^{\Phi} dr}.
\eq
 Then, in terms of $t$,  the above metric takes explicitly the canonical ADM form with the projectability
condition,
\bqn
\lb{metricB}
ds^{2} &=& - dt^{2} + \left(e^{\mu(r)} dt + e^{\nu(r)}dr\right)^{2} 
 + r^{2}d\Omega^{2},  ~~~
\eqn 
with
\bqn
\lb{coefficients}
\Phi(r) &=&  \nu(r) - \frac{1}{2}\ln\Big(1 -  e^{2\mu}\Big), \nb\\  
\Psi(r) &=& \frac{1}{2}\ln\Big(1 - e^{2\mu}\Big),\nb\\
u_{\mu} &=& \sqrt{1 - e^{2\mu}} \delta^{t}_{\mu}
- \frac{ e^{\mu + \nu}}{\sqrt{1 -  e^{2\mu}}} \delta^{r}_{\mu},\nb\\
r_{\mu} &=&\frac{e^{\nu}}{\sqrt{1 - e^{2\mu}}}  \delta^{r}_{\mu},\;\;\;
u^{\mu} = - \frac{1}{\sqrt{1 - e^{2\mu}}} \delta^{\mu}_{t}, \nb\\
r^{\mu} &=&\frac{e^{\mu}}{\sqrt{1 - e^{2\mu}}}  \delta^{\mu}_{t}
        + {e^{-\nu}}{\sqrt{1 - e^{2\mu}}}  \delta^{\mu}_{r}.
\eqn
Clearly, to have the coordinate transformations be real, we must assume
\bq
\lb{C.20a}
e^{2\Psi} \le 1.
\eq
$\Psi$ is often written as \cite{Wang}
\bq
\lb{C.20b}
e^{2\Psi} =1 - \frac{2m(r)}{r},
\eq
where $m(r)$ represents the gravitational mass within the shell $r$. When $m(r) \ge 0$, the condition (\ref{C.20a})
is satisfied identically. 

 It should be noted that the coordinate transformations given by Eq. (\ref{trans}) are not allowed by the 
foliation-preserving diffeomorphisms (\ref{gauge}). In particular, the action of Eq. (\ref{2.4}) is not invariant, because
now the extrinsic curvature and Ricci tensors $K_{ij}$ and $R_{ij}$  no longer  behave like tensors under
these transformations.

Let us note here that the definitions of the energy density $\rho_o$, the radial pressure $p_R$ and the heat flow $q_o$ are different from the ones ($\rho_H$, $p_r$, $q$) given by Eq. (\ref{3.3l}), which are defined   by assuming that the fluid
 is comoving with respect to the canonical ADM frame (\ref{metricB}). The relation between the two sets of quantities is the following
\bqn
\rho_H&=&{1 \over 1-e^{2\mu}} (\rho_o +e^{2\mu}p_R   -2 e^\mu q_o),\nb \\
p_r &=&{1 \over 1-e^{2\mu}} (p_R + e^{2\mu} \rho_o -2 e^\mu q_o), \nb \\
q&=&{1 \over 1-e^{2\mu}} \left[ (1+ e^{2\mu})q_o - e^\mu (\rho_o + p_R) \right].\eqn

The nonvanishing components of the Einstein tensor  for the metric (\ref{metricB}) are given by
\bqn
\lb{G00}
G_{00} &=& \frac{\left(1 - e^{2\mu}\right)e^{2(\mu -\nu)}}{r^{2}}\Big[2r\mu' 
    + \left(1 - e^{-2\mu}\right)\left(1 - 2r\nu'\right)\nb\\
    & & ~~~~~~~~~~~~~~~~~~~~~~~~~ + e^{2(\nu - \mu)}\Big],\nb\\
\lb{G01}
G_{01} &=& - \frac{e^{3\mu -\nu}}{r^{2}}\Big[2r\mu' 
    + \left(1 - e^{-2\mu}\right)\left(1 - 2r\nu'\right)\nb\\
    & & ~~~~~~~~~~~~~~~~~~~~~~~~~ + e^{2(\nu - \mu)}\Big],\nb\\    
\lb{G11}
G_{11} &=&  - \frac{e^{2\mu}}{r^{2}}\Big[2r\left(\mu' - \nu'\right)
    + \left(1 - e^{-2\mu}\right) + e^{2(\nu - \mu)}\Big],\nb\\  
  \lb{G22}
G_{22} &=&  - re^{2(\mu - \nu)}\Big[r\left(\mu'' + 2\mu'^{2} - \nu'\mu'\right)
    +2\mu' \nb\\
    & & ~~~~~~~~~~~~~~~~~~~~~~~~~ 
  - \left(1 - e^{-2\mu}\right)\nu'\Big].
\eqn 
Then, for an anisotropic fluid (\ref{EMT}), the Einstein field equations, $G_{\mu\nu} = 8 \pi G T_{\mu\nu}$,
yield,
\bqn
\lb{B.7a}
& &  2r\mu'  -  2r \left(1 - e^{-2\mu}\right)\nu' - \left(1 - e^{2\nu}\right)e^{-2\mu} + 1 \nb\\
& & ~~~~~~~~~~~~~~~ =  {8\pi G r^{2}e^{2(\nu - \mu)}}  \rho_o,\;\; [G_{00}], \\
\lb{B.7b}
& &   \left(1 - e^{-2\mu}\right)\nu' = -     4\pi G r e^{2(\nu - \mu)} \left(\rho_o 
                                              + p_{R} - 2e^{\mu} q_o\right),\nb\\
                                              & &  ~~~~~~~~~~~~~~~  [e^{2\nu}G_{00} + (1 - e^{2\mu})G_{11}], \\
\lb{B.7c}
& &  r\left(\mu'' + 2\mu'^{2} - \nu'\mu'\right) + 2\mu'  -  \left(1 - e^{-2\mu}\right) \nu'\nb\\
& &  ~~~~~~~~~~~~~~~ = - 8\pi G re^{2(\nu - \mu)}p_{\theta},         \;\; [G_{22}],\\
\lb{B.7d}
& &  q_o = 0,\;\;\; [e^{\mu + \nu}G_{00} + (1 - e^{2\mu})G_{01}].                    
\eqn                                              
The last equation shows clearly that  in GR heat flow along radial direction is not allowed in static spherically symmetric
spacetimes. 

The conservation laws $\nabla^{\nu}T_{\nu\mu} = 0$, on the other hand,  give
\bqn
\lb{B.8a1}
& & 2rq_o \mu' + \big(1 - e^{-2\mu}\big)\big(r q'_o + 2q_o\big) = 0,\\
\lb{B.8a2}
&& \Big[\big(\rho_o + p_{R}\big) - 2e^{\mu} q_o\Big]\mu' +  \big(1 -  e^{-2\mu}\big)\Big[\big(p_{R}' - e^{\mu}q'_o \big)\nb\\
& & ~~~~~~~~~ + \frac{2}{r}\big(p_{R} - p_{\theta} - e^{\mu} q_o \big)\Big] = 0. 
\eqn
For $q_o = 0$, Eq. (\ref{B.8a1}) is satisfied identically, while Eq. (\ref{B.8a2}) reduces to
\bq
\lb{B.8a}
\big(\rho_o + p_{R}\big)\mu' +  \big(1 -  e^{-2\mu}\big)\Big[p_{R}' + \frac{2}{r}\big(p_{R} - p_{\theta}\big)\Big] = 0,\;
(q_o = 0).~~
\eq

When $\nu = 0$, from Eq. (\ref{B.7b}) we see that $p_{R} = - \rho_o$. If in addition we have a perfect fluid $p_{R} = p_{\theta}$, from Eq. (\ref{B.8a}) we find that $p_{R}' = 0$, that is, the pressure is constant. Then, from Eq. (\ref{B.7a})
we find that $\mu$ is exactly given by Eq. (\ref{dS}), which is identically the de Sitter Schwarzshcild
solution written in the ADM form \cite{Visser}.   

When $\nu = 0$ and $p_{\theta} = \gamma p_{R}$, from Eqs. (\ref{B.7a}) -  (\ref{B.8a}) we find that
\bqn
\lb{B.9}
\mu &=&  \frac{1}{2}\ln\left[\frac{M}{r} + \left(\frac{r}{\ell}\right)^{2\gamma}\right] + \mu_{0},\nb\\
\rho_o &=& - p_{R} = -\gamma^{-1} p_{\theta} =  - c_{0}r^{2(\gamma - 1)},
\eqn
where $\mu_{0}$ is given by Eq.(\ref{3.10}), and $c_{0}$ is a constant. To have $\rho_o$ non-negative we must assume $c_{0} < 0$,
while finiteness at the center requires $\gamma > 1$.  As a result, we find that $\rho_o + p_{\theta} = (1-\gamma)\rho_o < 0$, that is,
the fluid does not satisfies the weak energy condition \cite{HE72}. In fact, it does not satisfies any of the three energy
conditions.

\section*{Appendix C:  Singular Behavior of Extrinsic Curvature $K$}
\renewcommand{\theequation}{C.\arabic{equation}} \setcounter{equation}{0}

For the static spacetimes described by metric (\ref{3.1b}), we have 
\bq
\lb{C.1}
K = e^{\mu-\nu}\left(\mu' + \frac{2}{r}\right).
\eq

For the (anti-) de Sitter Schwarzschild solutions (\ref{dS}), $K$ is given by \cite{CW}
\bq
\lb{C.2}
K =  \sqrt{\frac{3M + \Lambda r^{3}}{12 r^{3}}}
\left(4- \frac{3M -  2\Lambda r^{3}}{3M + \Lambda r^{3}}\right).
\eq
As noticed in \cite{CW}, $K$ is singular for the anti-de Sitter Schwarzschild solution ($\Lambda < 0$)  
not only at the center $r = 0$ but also at $r = (3M/|\Lambda|)^{1/3}$. The latter singularity is absent in GR.

For the Ricci-flat solutions given by Eq. (\ref{3.10}), we have
\bqn
\lb{C.3}
K &=&\frac{e^{\mu_{0}}}{2r^{3/2}\left[M + r \left(\frac{r}{\ell}\right)^{2\gamma}\right]^{1/2}}\Bigg\{3M\nb\\
& & + 2r(2+ \gamma )  \left(\frac{r}{\ell}\right)^{2\gamma}\Bigg\},
\eqn
which is singular at the center, unless $M=0$ and $\gamma \ge 1$.

For the solutions of Eq. (\ref{3.17}) with $b = 0$, we find that
\bq
\lb{C.4}
K  = 3a,
\eq
and therefore everywhere finite.

For the solutions of Eq. (\ref{3.33}), we find that
\bq
\lb{C.5}
K = 3 e^{\mu_0} \sqrt{c_2 a_2 a_3} (1+ {\cal{O}}(r)),
\eq
and therefore finite  at $r = 0$.

For the solutions of Eq. (\ref{3.45}), we find that
\bqn
\lb{C.6} 
K= 3e^{\mu_{0}} \sqrt{1 - \frac{k}{6}r^{2}} ,
\eqn
and therefore everywhere finite.

For the solutions of Eq. (\ref{3.51}), we find that as $r \to 0$
\bq
\lb{C.7} 
K \simeq {3 \over 2(1-\xi)} \left(a_{1}c_{1} + a_{2}c_{2}\right)^{\frac{1-2\xi}{2(1-\xi)}}r^{\frac{3(2\xi - 1)}{2(1-\xi)}},
\eq
with non-singular corrections. Thus, when $a_{1}c_{1} + a_{2}c_{2} \not= 0$ and $1/2 \le  \xi < 1$, $K$ is finite at the center. For $a_{1}c_{1} + a_{2}c_{2} = 0$ one obains the same condition for $\xi$, but these solutions are forbidden from the regularity condition  (\ref{3.53}).

We have not provided here the singular behavior of the invariant $K_{ij}K^{ij}$, but it turns out to have similar behavior as the one of the trace of the extrinsic curvature $K$.


\end{document}